\documentclass[sn-mathphys,pdflatex]{sn-jnl}% referee option is meant for double line spacing

\jyear{2021}%

\theoremstyle{thmstyleone}%
\theoremstyle{thmstyletwo}%

\theoremstyle{thmstylethree}%

\usepackage{graphicx}
\usepackage{xcolor}
\usepackage{enumitem}
\usepackage{float}

\newcommand{\ket}[1]{\vert #1 \rangle}
\newcommand{\bra}[1]{\langle #1 \vert}
\newcommand{\vn}{\mathbf{n}}
\newcommand{\vna}{\mathbf{n}_{\textsc{\scriptsize a}}}
\newcommand{\vnb}{\mathbf{n}_{\textsc{\scriptsize b}}}
\newcommand{\mb}{m_{\textsc{\scriptsize b}}}
\newcommand{\ma}{m_{\textsc{\scriptsize a}}}
\newcommand{\aver}[1]{\langle #1 \rangle}

\newcommand{\smtextsc}[1]{\textsc{\scriptsize #1}}

\newcommand{\eref}[1]{(\ref{#1})}
\graphicspath{{./figures/}}

\newcommand{\epsarg}[1]{\varepsilon_{\small\textsc{\scriptsize #1}}}

\usepackage{xparse,xcoffins}

\ExplSyntaxOn
\NewCoffin\imagecoffin
\NewCoffin\labelcoffin

\keys_define:nn { miguel/label }
 {
  label   .tl_set:N = \l_miguel_label_tl,
  labelbox .bool_set:N = \l_miguel_label_box_bool,
  labelbox .default:n = true,
  fontsize .tl_set:N = \l_miguel_label_size_tl,
  fontsize .initial:n = \footnotesize,
  pos .choice:,
  pos/nw .code:n = \tl_set:Nn \l_miguel_label_pos_tl { left,up },
  pos/ne .code:n = \tl_set:Nn \l_miguel_label_pos_tl { right,up },
  pos/sw .code:n = \tl_set:Nn \l_miguel_label_pos_tl { left,down },
  pos/se .code:n = \tl_set:Nn \l_miguel_label_pos_tl { right,down },
  pos/n .code:n = \tl_set:Nn \l_miguel_label_pos_tl { hc,up },
  pos/w .code:n = \tl_set:Nn \l_miguel_label_pos_tl { left,vc },
  pos/s .code:n = \tl_set:Nn \l_miguel_label_pos_tl { hc,down },
  pos/e .code:n = \tl_set:Nn \l_miguel_label_pos_tl { right,vc },
  pos .initial:n = nw,
  unknown .code:n   = \clist_put_right:Nx \l_miguel_label_clist
                       { \l_keys_key_tl = \exp_not:n { #1 } }
 }
\clist_new:N \l_miguel_label_clist
\box_new:N \l_miguel_label_box
\box_new:N \l_miguel_label_image_box

\NewDocumentCommand{\xincludegraphics}{O{}m}
 {
  \group_begin:
  \tl_clear:N \l_miguel_label_tl
  \clist_clear:N \l_miguel_label_clist
  \keys_set:nn { miguel/label } { #1 }
  \tl_if_empty:NTF \l_miguel_label_tl
   {
    \miguel_includegraphics:Vn \l_miguel_label_clist { #2 }
   }
   {
    \SetHorizontalCoffin\imagecoffin
     {
      \miguel_includegraphics:Vn \l_miguel_label_clist { #2 }
     }
    \SetHorizontalCoffin\labelcoffin
     {
      \raisebox{\depth}
       {
        \bool_if:NTF \l_miguel_label_box_bool
         { \fcolorbox{white}{white}{\l_miguel_label_size_tl\l_miguel_label_tl} }
         { \l_miguel_label_size_tl\l_miguel_label_tl }
       }
     }
    \SetVerticalPole\imagecoffin{left}{-5pt+\CoffinWidth\labelcoffin/2}
    \SetVerticalPole\imagecoffin{right}{\Width-3pt-\CoffinWidth\labelcoffin/2}
    \SetHorizontalPole\imagecoffin{up}{\Height-3pt-\CoffinHeight\labelcoffin/2}
    \SetHorizontalPole\imagecoffin{down}{-5pt+\CoffinHeight\labelcoffin/2}
    \use:x{\JoinCoffins\imagecoffin[\l_miguel_label_pos_tl]\labelcoffin[vc,hc]} 
    \TypesetCoffin\imagecoffin
   }
   \group_end:
 }
\NewDocumentCommand{\setlabel}{m}
 {
  \keys_set:nn { miguel/label } { #1 }
 }

\cs_new_protected:Nn \miguel_includegraphics:nn
 {
  \includegraphics[#1]{#2}
 }
\cs_generate_variant:Nn \miguel_includegraphics:nn { V }

\ExplSyntaxOff

\begin{document}

\title{Tomographic entanglement indicators from NMR experiments}


\author*[1,4]{\fnm{B.} \sur{Sharmila}}\email{Sharmila.Balamurugan@warwick.ac.uk}

\author[2,3]{\fnm{V.~R.} \sur{Krithika}}

\author[2,3]{\fnm{Soham} \sur{Pal}}

\author[2,3]{\fnm{T.~S.} \sur{Mahesh}}

\author[1]{\fnm{S.} \sur{Lakshmibala}}

\author[1]{\fnm{V.} \sur{Balakrishnan}}

\affil*[1]{\orgdiv{Department of Physics}, \orgname{Indian Institute of Technology Madras}, \orgaddress{\city{Chennai} \postcode{600036}, \country{India}}}

\affil[2]{\orgdiv{Department of Physics}, \orgname{Indian Institute of Science Education and Research}, \orgaddress{\city{Pune} \postcode{411008}, \country{India}}}

\affil[3]{\orgdiv{NMR Research Center}, \orgname{Indian Institute of Science Education and Research}, \orgaddress{\city{Pune} \postcode{411008}, \country{India}}}

\affil[4]{\orgdiv{Present address: Department of Physics}, \orgname{University of Warwick}, \orgaddress{\city{Coventry} \postcode{CV4 7AL}, \country{UK}}}

%%==================================%%

\abstract{In recent years, the performance of different entanglement indicators obtained directly from tomograms has been assessed in continuous-variable and hybrid quantum systems. In this paper, we carry out this task in the case of spin systems. We compute the entanglement indicators from actual experimental data obtained from three liquid-state NMR experiments, and compare them with standard entanglement measures calculated  from  the corresponding density matrices, both experimentally reconstructed and numerically computed. The gross features of entanglement dynamics and spin squeezing properties are found to be reproduced  by these entanglement indicators. However, the extent to which these indicators and spin squeezing  track  the entanglement during time evolution of the multipartite systems in the NMR experiments  is very sensitive to the precise nature and strength of interactions as well as the manner in which the full system is partitioned into subsystems. We also use the IBM quantum computer to implement equivalent circuits that capture the dynamics of the multipartite system in one of the NMR experiments. We compute and compare the entanglement indicators obtained from the tomograms corresponding to the experimental execution and simulation of these equivalent circuits. This exercise  shows that these indicators can estimate the degree of entanglement without necessitating detailed state reconstruction procedures, establishing the advantage of the tomographic approach.}


\maketitle

	\section{\label{sec:intro}Introduction}
	
The outcome of the  measurement of a
physical quantity   in  a quantum system 
is essentially  a histogram of the state of the system  in terms of the 
 basis kets corresponding to the 
 observable concerned. 
 A tomogram comprises a set of such histograms, 
 obtained by informationally complete measurements of a quorum of  
 observables. The deduction of the density 
 matrix of the system from tomograms 
 is the primary objective of  quantum state reconstruction.  This reconstruction of the state 
 from tomograms is not error-free, as it entails 
 statistical techniques~\cite{photonnumbertomo2}. 
 This drawback is present even in as simple a
 situation as a 2-qubit bipartite system, and is 
 naturally more pronounced when there is   multipartite entanglement~\cite{QubitRecon2016}.
The difficulties involved in developing  scalable reconstruction procedures for cases involving a large number of qubits 
 have been described in~\cite{ReconTech2013,ReconTech2018}.
  
The data-processing resources needed  to obtain the quantum state from tomograms are known to increase with the number of qubits. In order to reconstruct an $n$-qubit density matrix using standard procedures, one needs large sets of positive-operator-valued measurements. This can be traced back to the large dimensions of the Hilbert space. Since standard reconstruction methods are very time consuming, other procedures involving compressed sensing~\cite{ReconCompSens1,ReconCompSens2,ReconCompSens3,ReconCompSens4} have been suggested, to characterize an unknown quantum state using only a subset of the data. While specific entangled multiqubit pure states with special structures in the density matrix have been reconstructed reasonably well, in general reconstruction of an arbitrary multiqubit state still poses serious challenges.

  In the light of the foregoing, it becomes 
  relevant to elicit  as much  information about the state as possible directly from the tomogram, 
  circumventing   
  the need to reconstruct the state.  In bipartite qubit systems this may be done~\cite{HuiKhoon} by using 
  the tomogram to compute the fidelity of the state 
 with respect to some specific target state, and then  
 comparing the accuracy of this procedure with that 
corresponding to reconstruction of the state. 
Of special interest  is the estimation 
 of entanglement from easily carried out 
 operations on the tomograms concerned.

 As is well known, 
quantum information processing relies 
heavily on entanglement. 
Model systems have revealed the occurrence of 
 both abrupt emergence and sudden death of entanglement~\cite{eberly}, as well as the 
 `collapse' of the entanglement 
 to a constant non-zero value for 
 sizable time intervals~\cite{pradip1}. 
	
The basic measure of the entanglement between the two subsystems A and B of a bipartite system AB is the subsystem von Neumann entropy 
(SVNE) 
$\xi_{\smtextsc{svne}}=-\mathrm{Tr}\,(\rho_{i} \,\log_{2} \,\rho_{i})$,  where $\rho_{i}$ ($i=\mathrm{A,\,B}$) is the reduced (or subsystem)  density matrix. 
When A and B are subsystems of a 
 multipartite system, the mutual information
	\begin{equation}
	\xi_{\smtextsc{qmi}}=\xi_{\smtextsc{svne}}^{(\smtextsc{a})}+ \xi_{\smtextsc{svne}}^{(\smtextsc{b})}- \xi_{\smtextsc{svne}}^{(\smtextsc{ab})}.
	\label{eqn:qmi}
	\end{equation}
	is a convenient measure of the correlations between A and B. Another standard measure of quantum correlations is provided by 
	the quantum  discord 
	$D(\mathrm{B}:\mathrm{A})$ between the two subsystems A and B. If projective measurements are carried out on A, this discord  is defined as
	\begin{equation}
	D(\mathrm{B}:\mathrm{A}) = \xi_{\smtextsc{svne}}^{(\smtextsc{a})}-\xi_{\smtextsc{svne}}^{(\smtextsc{ab})} + 
	\mathrm{min}_{\lbrace \mathcal{O}_{i}^{\smtextsc{a}} \rbrace}\Big\{-\sum_{j} p^{\smtextsc{a}}_{ij} \text{Tr} \left(\varrho_{ij} \log_{2} \varrho_{ij}\right)\Big\},
	\label{eqn:discord}
	\end{equation}
	where $\lbrace \mathcal{O}_{i}^{\smtextsc{a}} \rbrace$ is a set of subsystem observables 
	pertaining  to A. Here, 
\begin{equation}
p^{\smtextsc{a}}_{ij}=\text{Tr} [(\Pi_{ij}^{\smtextsc{a}} \otimes \mathbb{I}_{\smtextsc{b}}) \rho_{\smtextsc{ab}}]
\end{equation}
and 
	\begin{equation}
	\varrho_{ij}=\frac{\text{Tr}_{\smtextsc{a}} [(\Pi_{ij}^{\smtextsc{a}} \otimes \mathbb{I}_{\smtextsc{b}}) \rho_{\smtextsc{ab}} (\Pi_{ij}^{\smtextsc{a}} \otimes \mathbb{I}_{\smtextsc{b}})]}{p^{\smtextsc{a}}_{ij}},
	\end{equation}
where $\lbrace\Pi_{ij}^{\smtextsc{a}} \rbrace$ is the set of projection operators corresponding to $\mathcal{O}_{i}^{\smtextsc{a}}$ and $\mathbb{I}_{\smtextsc{b}}$ denotes the identity operator in B. $D(\mathrm{A}:\mathrm{B})$ is similarly defined 
when projective measurements are carried out on B. In general, $D(\mathrm{A}:\mathrm{B})\neq D(\mathrm{B}:\mathrm{A})$.
	
The computation of any of the  measures   
$\xi_{\smtextsc{svne}}$, $\xi_{\smtextsc{qmi}}$ and 
$D$ entails  prior  knowledge of the density matrix of the full system. 
It would therefore be expedient to deploy indicators of entanglement based directly on tomograms. This has been done  
in the case of  continuous-variable (CV) systems involving a radiation field, illustrating  
the usefulness of this approach.  For instance, the entanglement between the radiation fields in the output ports of a quantum beamsplitter has  been identified 
qualitatively from tomograms~\cite{sudhrohithbs}. 
Further, 
a quantitative analysis of entanglement using 
several indicators based  directly on tomograms in bipartite CV and multipartite hybrid quantum (HQ) systems has been carried out ~\cite{sharmila2,sharmila4}, and the results 
have been compared with  standard measures of entanglement such as $\xi_{\smtextsc{qmi}}$.

	In this paper, we present  such a comparison for spin systems. The exercise is important in order to understand the role and usefulness  of tomographic entanglement indicators in the broad framework comprising CV, spin, and HQ systems. A major difference compared to earlier work in this regard is that the investigations reported here are based on actual  experimental data from three different liquid-state NMR experiments, rather than on numerically generated states alone. These experiments are labelled I, II, and III in subsequent sections. The details regarding the methods of state preparation, the data obtained, and the errors involved in the experiment have been reported elsewhere~\cite{nmrExpt,nmrExpt1}. These NMR experiments furnish a good platform for testing the efficacy of our indicators. NMR techniques are commonly used in quantum information processing (see, for instance, \cite{NMRtomography}) and extensive experimental investigations in NMR have demonstrated spin-squeezing (see, for instance, \cite{2003nmr,2015nmr}), generation (see, for instance,~\cite{1998nmr}) and estimation(see, for instance,~\cite{2002nmr,2012nmr}) of entanglement. Experiments I and III have been performed on $\vphantom{}^{13}\mathrm{C}$, $\vphantom{}^{1}\mathrm{H}$ and $\vphantom{}^{19}\mathrm{F}$ spin-half nuclei in dibromofluoromethane (DBFM) dissolved in deuterated acetone~\cite{nmrExpt}. NMR experiment II has been performed on $\vphantom{}^{19}\mathrm{F}$ and $\vphantom{}^{31}\mathrm{P}$ spin-half nuclei in sodium flourophosphate (NaFP) dissolved in $\mathrm{D}_{2}\mathrm{O}$~\cite{nmrExpt1}. For our present purposes, only the tomograms obtained in these experiments are of direct relevance. 

In all these experiments, due to the smallness of the Hilbert space, the relevant density matrices have been easily reconstructed from the tomograms through a two-step procedure. The deviation matrix is first reconstructed from the tomogram, and subsequently the density matrix is readily computed from the deviation matrix. This procedure is given in~\cite{nmrExpt}. With increase in the dimension of the Hilbert space, this program is not straightforward, and in fact poses several challenges as mentioned earlier. In this paper, we examine if this two-step program can be circumvented, and squeezing and entanglement properties of the system can be directly read off from the tomograms. For this purpose, in subsequent sections, the entanglement indicators obtained solely from the tomograms are compared with entanglement measures obtained from the density matrices in all the three experiments. We have verified that the entanglement indicators obtained through both these procedures are in good agreement with each other. This result therefore establishes that the tomographic approach is a powerful alternative in estimating nonclassical effects. By extrapolation, it is evident that this approach is potentially very useful for systems with large Hilbert spaces, be they an array of interacting qubits or coupled CV systems.

In this paper, we have also addressed the problem of identifying the optimal data set (equivalently, the number of measurements) needed to obtain good entanglement indicators. This amounts to identifying the optimal number of tomographic `slices' needed for this purpose. Our investigations in this regard have been made using data from the three NMR experiments mentioned above, and we have summarised our observations.
	
	The plan of this paper is as follows. In Sec.~\ref{sec:revIndics}, we define spin tomograms, recapitulate  their salient features, and describe the entanglement indicators defined in terms of these tomograms. In Sec.~\ref{sec:squeeze_prop}, we briefly review the procedure for obtaining  squeezing properties both from the tomograms directly and from the state. In Sec.~\ref{sec:NMRexpt}, we analyse tomograms corresponding to NMR experiment I. We also comment on the spin, higher-order, and entropic squeezing properties deduced from tomograms for this system. In Sec.~\ref{sec:NMRexpt2}, we examine NMR experiments II and III, and analyse the squeezing properties and the entanglement indicators obtained directly from relevant tomograms. The three NMR experiments considered here are on either bipartite or tripartite spin systems, and bipartite entanglement has been investigated in all cases. Squeezing properties have been examined for the full systems. We conclude with some brief additional remarks. In the Appendix, we have proposed equivalent circuits for the NMR experiment I at specific instants of time,  and implemented them in the IBM quantum computing platform (IBM Q). Corresponding tomograms have been obtained both from the experimental runs in the IBM Q platform, and from numerical simulations using the IBM open quantum assembly language (QASM) simulator~\cite{ibm_main,qiskit}. The idea behind using the IBM platform is two-fold: (i) To implement an equivalent circuit that models the dynamics in the system of interest, and (ii) to demonstrate that the extent of entanglement can be assessed in this equivalent setup without resorting to detailed state reconstruction.
	
\section{\label{sec:revIndics}Spin tomograms}
	
\subsection{Notation and definitions}
	
The quorum of observables ~\cite{thew} for a single 
spin-$\frac{1}{2}$  system is given by
\begin{align}
\label{eqn:atomops}
\nonumber \sigma_{x}=\frac{1}{2}
 (\ket{\uparrow}\bra{\downarrow}&+\ket{\downarrow}\bra{\uparrow}),
\;\sigma_{y}=\frac{1}{2} i(\ket{\downarrow}\bra{\uparrow}-\ket{\uparrow}\bra{\downarrow}), \\
& \sigma_{z}=\frac{1}{2} (\ket{\uparrow}\bra{\uparrow}-\ket{\downarrow}\bra{\downarrow}).
\end{align}

	$\ket{\downarrow}$ and $\ket{\uparrow}$ 
	stand for  the 
	eigenstates of $\sigma_{z}$, as usual.
	For notational convenience in this section, we denote $\ket{\downarrow}$ and $\ket{\uparrow}$ by $\ket{-\frac{1}{2}}$ and $\ket{+\frac{1}{2}}$ respectively, so that  $\sigma_{z}\ket{m}=m\ket{m}$  where $m=\pm 1/2$. The state  $\ket{\vartheta,\varphi,m} = U(\vartheta,\varphi) \ket{m}$,   
	where 
	\begin{equation}
	\nonumber U(\vartheta,\varphi)=\begin{bmatrix}
		\cos \left(\frac{\vartheta}{2}\right) e^{i \varphi/2} & \sin \left(\frac{\vartheta}{2}\right) e^{i \varphi/2} \\
		-\sin \left(\frac{\vartheta}{2}\right) e^{-i \varphi/2} & \cos \left(\frac{\vartheta}{2}\right) e^{-i \varphi/2} \end{bmatrix}.
	\end{equation}
Using the unit vector $\vn$ to denote the 
direction specified by $(\vartheta, \varphi)$, 
the  qubit tomogram is  given by 
	\begin{equation}
	\label{eqn:spintomogram}
	w(\vn,m)=\bra{\vn,m}\rho_{\smtextsc{s}}\ket{\vn,m}
	\end{equation}
	where $\rho_{\smtextsc{s}}$ is the qubit density matrix. There exists a basis for  each value of $\vn$, so that $\sum\limits_{m} w(\vn,m)=1$. 
	
	It is straightforward to  extend the foregoing to multipartite spin systems. For a bipartite system, 
	in particular,  the quorum is the set of 9 
	direct product combinations of the operators defined in Eq.~\eref{eqn:atomops}. The bipartite tomogram
	\begin{equation}
	w(\vna,\ma;\vnb,\mb)= \aver{\vna,\ma;\vnb,\mb \vert \rho_{\smtextsc{ab}}\vert\vna,\ma;\vnb,\mb},
	\label{eqn:2modeTomoDefn}
	\end{equation}
	where $\rho_{\smtextsc{ab}}$ is the bipartite density matrix,  $\sigma_{i z} \ket{m_{i}} = m_{i} \ket{m_{i}}$ ($i=\mathrm{A,B}$), and  
$\ket{\vna,\ma;\vnb,\mb}$ stands for 	
	$\ket{\vna,\ma} \otimes \ket{\vnb,\mb}$. 
	The normalisation condition is given by 
	\begin{equation}
	\sum\limits_{\ma} \sum\limits_{\mb} w(\vna,\ma;\vnb,\mb)=1
	\label{eqn:tomoNorm}
	\end{equation}
	for each $\vna$ and $\vnb$. 
	The reduced tomograms for  the 
	subsystems A and B are 
	\begin{equation}
	w_{\smtextsc{a}}(\vna,\ma)= \aver{\vna,\ma\vert\rho_{\smtextsc{a}}\vert\vna,\ma} =  \sum\limits_{\mb} w(\vna,\ma;\vnb,\mb),
	\label{eqn:tomo_2_mode_sub_A}
	\end{equation}
	for any fixed value of $\vnb$, and
	\begin{equation}
	w_{\smtextsc{b}}(\vnb,\mb)= \aver{\vnb,\mb\vert\rho_{\smtextsc{b}}\vert\vnb,\mb} = \sum\limits_{\ma} w(\vna,\ma;\vnb,\mb)
	\label{eqn:tomo_2_mode_sub_B}
	\end{equation}
	for any fixed value of $\vna$. The reduced density matrices $\rho_{\smtextsc{a}}$ and 
	$\rho_{\smtextsc{b}}$ are  given by $\mathrm{Tr}_{\smtextsc{b}}(\rho_{\smtextsc{ab}})$ 
	and  $\mathrm{Tr}_{\smtextsc{a}}(\rho_{\smtextsc{ab}})$. 
	
\begin{figure*}[t]
\centering
\includegraphics[width=0.4\textwidth]{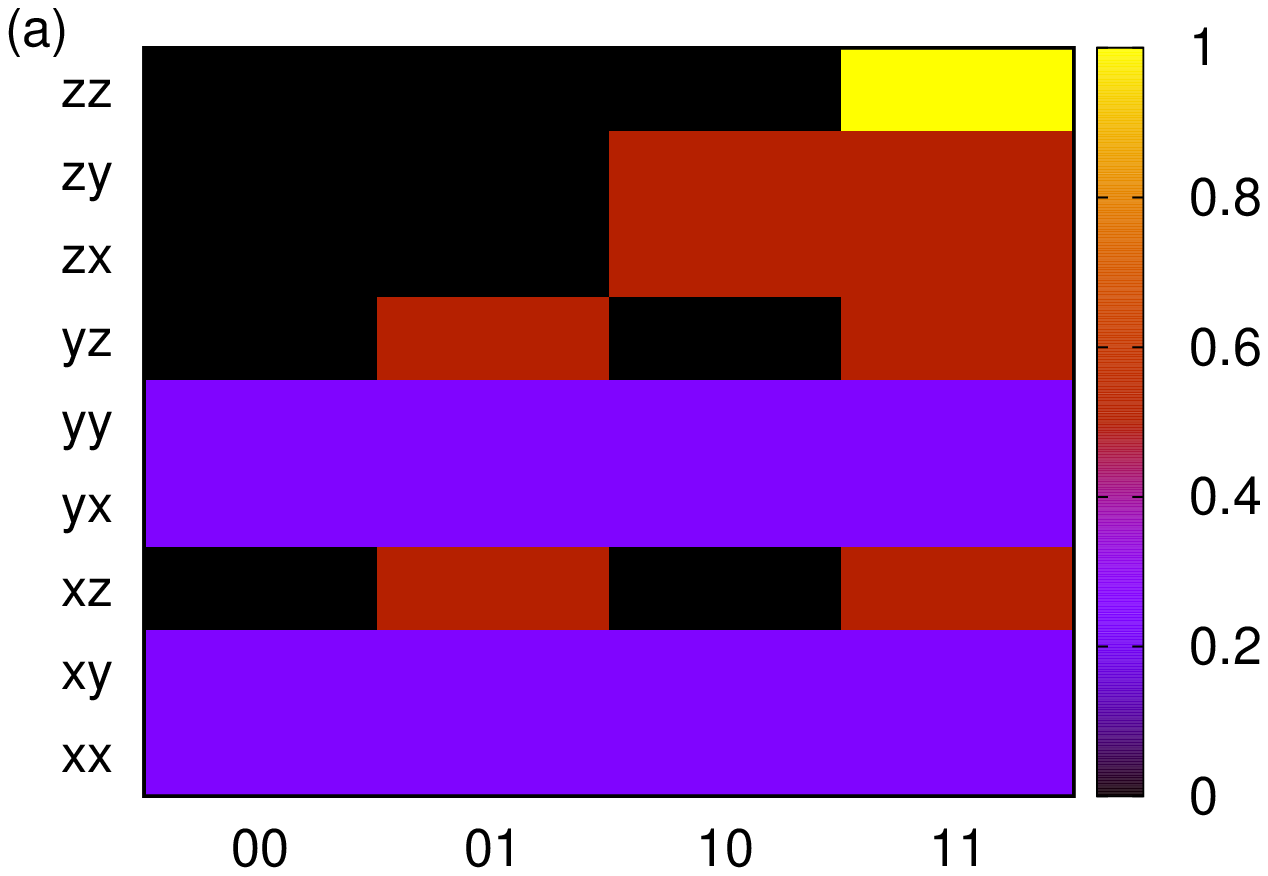}
\includegraphics[width=0.4\textwidth]{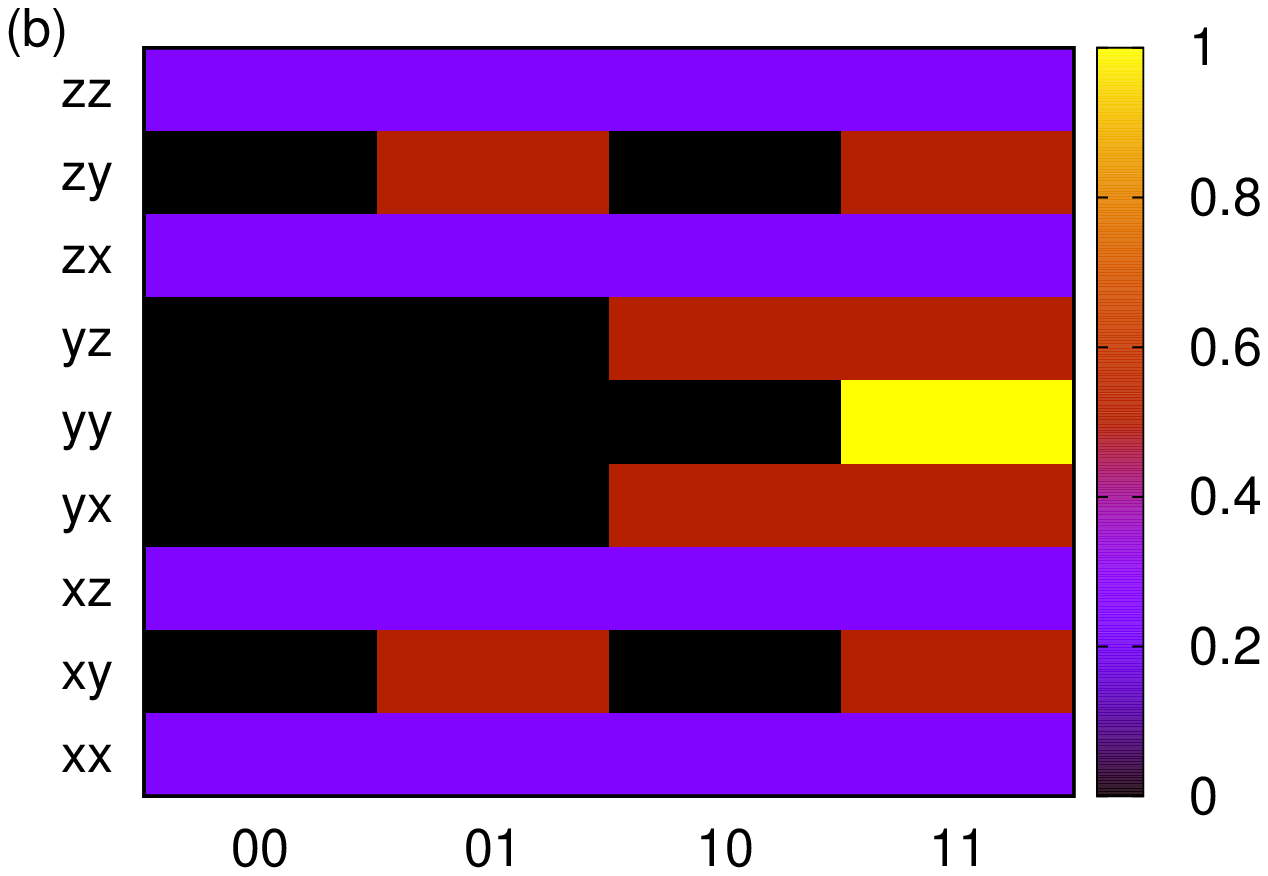}
\includegraphics[width=0.4\textwidth]{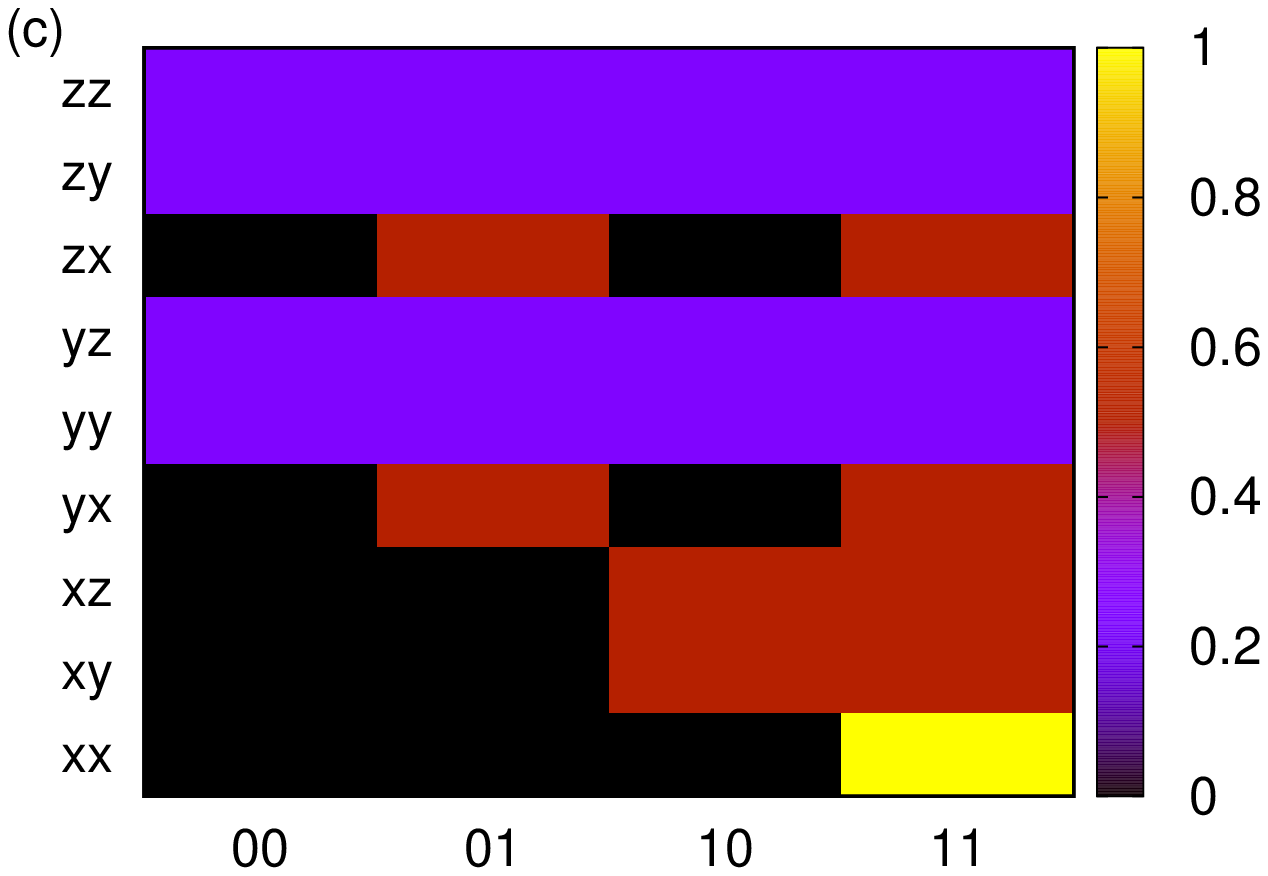}
\includegraphics[width=0.4\textwidth]{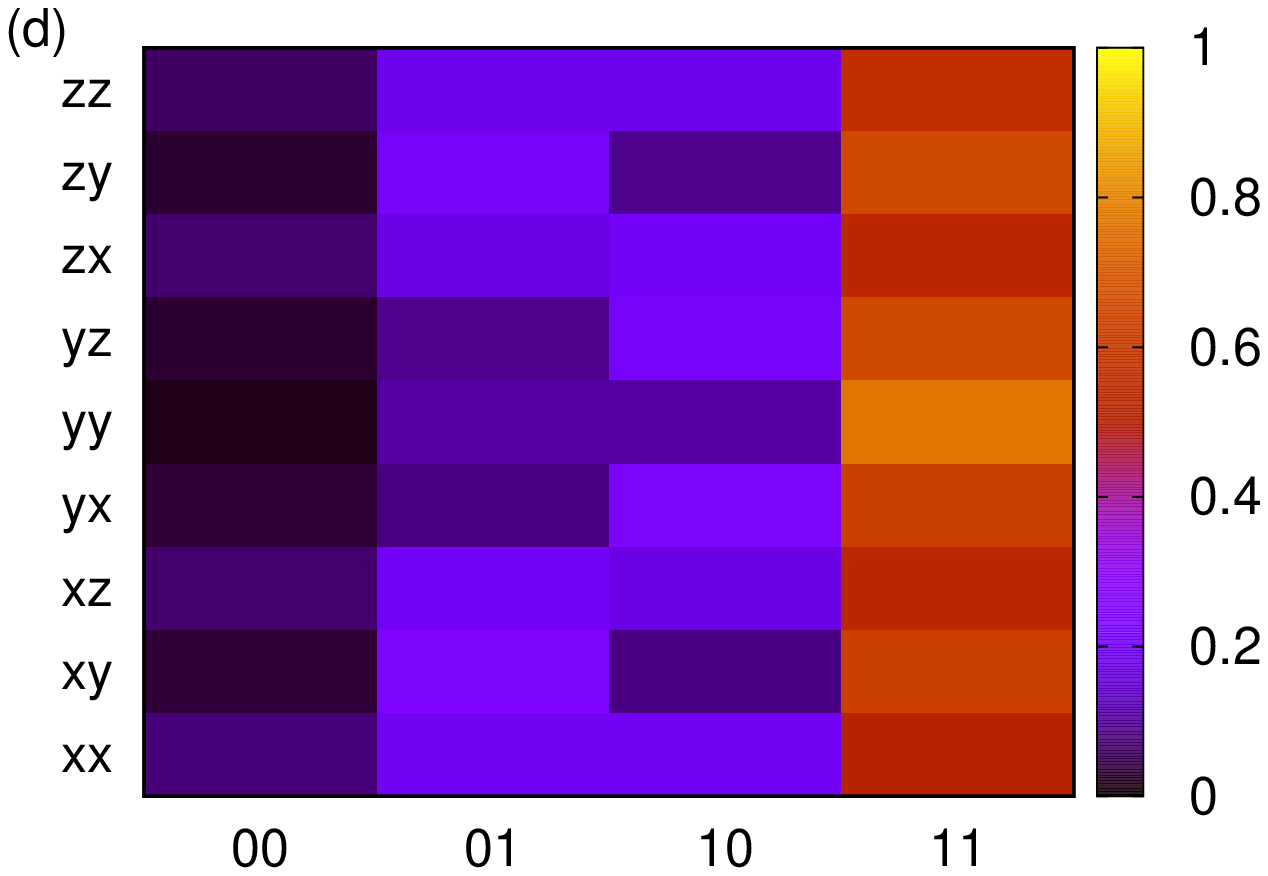}
\caption{Tomograms of spin coherent states with $(\theta, \phi)$ equal to (a) $(0,0)$, (b) $(\pi/2, \pi/2)$, (c) $(\pi/2, 0)$, and (d) $(\pi/3, \pi/3)$. The bases are denoted by $x$, $y$ and $z$ and the outcomes by $0$ and $1$.}
\label{fig:spin_CS_tomo}
\end{figure*}

It is convenient to plot  bipartite spin tomograms 
in the form of color maps. On the vertical axis, we 
mark the $9$ possibilities $xx$, $xy$, etc., in which the first and second labels correspond to subsystems A and 
B respectively. Each basis has four outcomes, namely, $00$, $01$, $10$, and $11$ where  $0$ and $1$  refer  to $m=-1/2$ and $m=+1/2$ respectively. Hence each color map is essentially a representation of a 
$(9\times 4)$ matrix. 
As a specific example, we present 
the spin tomograms for a bipartite spin coherent state
parametrized by angles $\theta, \phi$
(where $0\leqslant\theta\leqslant\pi, \;0\leqslant \phi < 2 \pi$), 
given by 
\begin{eqnarray}
\nonumber\left[\cos\left(\theta/2\right) \ket{\uparrow}_{\textsc{a}} + e^{i \phi}\,\sin \left(\theta/2\right)  \ket{\downarrow}_{\textsc{a}}\right] \qquad \qquad\\
\otimes \left[\cos\left(\theta/2\right) \ket{\uparrow}_{\textsc{b}} +e^{i \phi}\, \sin\left(\theta/2\right)  \ket{\downarrow}_{\textsc{b}}\right].
\end{eqnarray} 
Figures \ref{fig:spin_CS_tomo} (a)-(d) show the 
tomograms for four different values of 
$\theta$ and  $\phi$.

\subsection{Tomographic entanglement indicators}

We  use tomographic entropies 	
to estimate the degree of  correlation between the subsystems.
The bipartite tomographic entropy is given by 
\begin{align}
\nonumber S(\vna,\vnb) = - \sum\limits_{\ma} &\sum\limits_{\mb} w(\vna,\ma;\vnb,\mb)\\
& \log_{2} \, w(\vna,\ma;\vnb,\mb).
\label{eqn:2modeEntropy}
\end{align}
For a given subsystem, the  tomographic entropy is
	\begin{equation}
	S(\vn_{i}) = - \sum\limits_{m_{i}} w_{i}(\vn_{i},m_{i}) \log_{2} \,[w_{i}(\vn_{i},m_{i})] \;\; (i= \smtextsc{A,B}).
	\label{eqn:1modeEntropy}
	\end{equation}
	Some of the correlators to be 
	defined are obtained from a section of the tomogram
	corresponding to specific values of $\vna$ and $\vnb$. The degree to which  such a correlator is 
	a satisfactory estimator of entanglement therefore depends on  the specific tomographic section involved. The correlators and the associated  entanglement indicators are defined as follows.

The first of the tomographic entanglement indicators we consider is the mutual information 
	$\epsarg{tei}(\vna,\vnb)$,   
	expressed in terms of  the  tomographic entropies as 
	\begin{equation}
	\epsarg{tei}(\vna,\vnb)=S(\vna) + S(\vnb) - S(\vna,\vnb).
	\label{eqn:epsTEI}
	\end{equation}
	Indicators based on the inverse participation ratio (IPR) are also found to be satisfactory estimators of entanglement~\cite{sharmila2,ViolaBrown}. The participation ratio is a measure of delocalisation in a given basis. The IPR corresponding to a bipartite system in the basis of the spin operators  is defined as
	\begin{equation}
	\eta_{\smtextsc{ab}}(\vna,\vnb) = \sum\limits_{\ma} \sum\limits_{\mb} [w(\vna,\ma;\vnb,\mb)]^{2}.
	\label{eqn:IPRab}
	\end{equation}
	The IPR for each subsystem is  
	\begin{equation}
	\eta_{i}(\vn_{i}) = \sum\limits_{m_{i}} [w_{i}(\vn_{i},m_{i})]^{2} \;\; 
	(i= \smtextsc{A,B}).
	\label{eqn:IPRi}
	\end{equation}
	The associated entanglement indicator  is given by 
	\begin{equation}
	\epsarg{ipr}(\vna,\vnb)
	=1+\eta_{\smtextsc{ab}}(\vna,\vnb)-\eta_{\smtextsc{a}}(\vna)-\eta_{\smtextsc{b}}(\vnb). 
	\label{eqn:epsIPR}
	\end{equation}  
Apart from $\epsarg{tei}(\vna,\vnb)$ and 
$\epsarg{ipr}(\vna,\vnb)$, 
we have examined two  
	other correlators that  are familiar in the context of classical tomograms. Given 
	two random variables $X$ and $Y$,   
	their Pearson correlation coefficient
	is defined as~\cite{SmithMIcorr}   
	\begin{equation}
	\textsc{PCC}(X,Y)
	= \frac{\mathrm{Cov}(X,Y)}{(\Delta X) (\Delta Y)}.
	\label{eqn:PCC}
	\end{equation}
where $\Delta X, \, \Delta Y$ are the standard deviations of $X$ and $Y$, and  
	$\mathrm{Cov}(X,Y)$ is their covariance.
In the present context, the correlation coefficient  
	$\textsc{PCC}(\ma,\mb)$ 
	calculated  for  fixed values of $\vna$ and $\vnb$ 
	is relevant. A simple definition of the 
	entanglement indicator is then 
	\begin{equation}
	\epsarg{pcc}(\vna,\vnb)=
	\vert \textsc{PCC}(\ma,\mb)\vert.
	\label{eqn:epsPCC}
	\end{equation}
	The modulus ensures that the 
	quantifier of entanglement is non-negative.  
It is to be noted that  $\textsc{PCC}(\ma,\mb)$ has an implicit dependence on the directions $\vna$ 
and $\vnb$ chosen, and that 
$\epsarg{pcc}(\vna,\vnb)$ 
captures the effect of linear correlations. 
	
	The second indicator $\epsarg{bd}$ that we shall use originates as 
	follows. Consider two discrete random variables 
	$X$ and $Y$, with a joint probability 
	distribution  $p_{XY}(x_{i},y_{j})$ and 
	marginal distributions $p_{X}(x_{i})= \sum_{j} p_{XY}(x_{i},y_{j})$ and 
	$p_{Y}(y_{j}) = \sum_{i} p_{XY}(x_{i},y_{j})$. 
	Their mutual information is the      
  Kullback-Leibler divergence~\cite{kullback,ITCoverThomas,MI_KL_link} 	
\begin{equation}
	D_{\smtextsc{kl}}[p_{XY}\!:\!p_{X}p_{Y}] 
	= \sum\limits_{i,j} \,p_{XY}(x_{i},y_{j})\,\log_{2}
	\frac{p_{XY}(x_{i},y_{j})}{p_{X}(x_{i}) p_{Y}(y_{j})},
	\label{eqn:KLmutualinfo}
	\end{equation}
The indicator   $\epsarg{tei}(\vna,\vnb)$ 
	defined in Eq.~\eref{eqn:epsTEI} 
	is just  the mutual information in the case of 
	spin tomograms, i.e.,  
	\begin{equation}
	\epsarg{tei}(\vna,\vnb)= 
	D_{\smtextsc{kl}}
	\big[w(\vna,\ma;\vnb,\mb)\!:\! w_{\smtextsc{a}}(\vna,\ma) w_{\smtextsc{b}}(\vnb,\mb)\big]. 
	\label{eqn:KL_MI_link}
	\end{equation}
	A simpler  alternative to the 
	Kullback-Leibler divergence is 
	the Bhattacharyya distance~\cite{KL_BD_link}
	\begin{equation}
	D_{\smtextsc{b}}[p_{XY}\!:\!p_{X}p_{Y}] =  
	-\log_{2}\,\Big\{\sum\limits_{i,j} \,\big[p_{XY}(x_{i},y_{j}) 
	p_{X}(x_{i}) p_{Y}(y_{j})\big]^{1/2}\Big\}. 
	\label{eqn:Bhattmutualinfo}
	\end{equation}
	An application of Jensen's inequality leads to the 
	bound
		$D_{\smtextsc{b}} \leqslant \frac{1}{2} D_{\smtextsc{kl}}$, so that  
	$D_{\smtextsc{b}}$ is an approximate (under-)estimate 
		of the mutual information. It is thus natural to 
		define, in analogy with Eq.~\eref{eqn:KL_MI_link}, 
		the entanglement indicator 
	\begin{equation}
	\epsarg{bd}(\vna,\vnb)= D_{\smtextsc{b}}[w(\vna,\ma;\vnb,\mb)\!:\! w_{\smtextsc{a}}(\vna,\ma) w_{\smtextsc{b}}(\vnb,\mb)].
	\label{eqn:epsBD}
	\end{equation}
We may further remove the  dependence on 
	($\vna$, $\vnb$) of the 
	indicators  defined above,  by  
	averaging over the 9 possible values corresponding to the 3 orthogonal vectors for each of the $\vn_{i}$ ($i=\mathrm{A,B}$). The average values of 
	$\epsarg{tei}, \, \epsarg{ipr}, \, \epsarg{pcc},$ and 
	$\epsarg{bd}$ thus obtained 
	will be denoted by $\xi_{\smtextsc{tei}}, \, \xi_{\smtextsc{ipr}}, \, \xi_{\smtextsc{pcc}},$ and   
	$\xi_{\smtextsc{bd}}$ respectively. 
	
	\section{\label{sec:squeeze_prop}Squeezing properties}
	
	In generic multipartite spin systems in which  the subsystem states are entangled, spin squeezing 
	is related to the  entanglement~\cite{SpinSqRev}. Bearing  this in mind, we assess the squeezing properties of the NMR experiments I, II and III via two different procedures. The first procedure is outlined in~\cite{kitagawa}. 
The  second method deduces squeezing properties directly from  the corresponding spin tomograms. We have verified that the results  obtained by two procedures are in agreement with each other, 
showing  the usefulness of tomograms in extracting squeezing properties. 
	
	We first summarize the salient features of the procedure used in~\cite{kitagawa}. This relies on the fact that the state of a system consisting of $N$ 
	spin-$\frac{1}{2}$ subsystems is squeezed if one of the components normal to the mean spin vector of the system has a variance less than $N/4$, 
which is the variance of the 
	corresponding spin coherent state. For instance, if $N=2$, the mean spin direction is given by 
	$\mathbf{v}_{\smtextsc{s}}(t) = \aver{\mathbf{J}_{2}(t)}/\vert\aver{\mathbf{J}_{2}(t)}\vert$ where 
	\begin{equation}
	\mathbf{J}_{2}= (\sigma_{\smtextsc{a} x} + \sigma_{\smtextsc{b} x}) \mathbf{e_{x}} + (\sigma_{\smtextsc{a} y} + \sigma_{\smtextsc{b} y}) \mathbf{e_{y}} + (\sigma_{\smtextsc{a} z} + \sigma_{\smtextsc{b} z}) \mathbf{e_{z}}.
	\label{eqn:spinJ}
	\end{equation}
Here $\aver{\mathcal{O}(t)}=\mathrm{Tr}_{\smtextsc{ab}}(\rho_{\smtextsc{ab}}(t)\mathcal{O})$ for any operator 
$\mathcal{O}$, where  $\rho_{\smtextsc{ab}}(t)$ is  the density matrix at time $t$. As a first step, the variance 
	\begin{equation}
	(\Delta \; \mathbf{J}_{2}\mathbf{\cdot}\mathbf{v}_{\perp})^{2} =  \aver{(\mathbf{J}_{2}\mathbf{\cdot}\mathbf{v}_{\perp})^{2}}
	\label{eqn:J2Vperp}
	\end{equation}
	is calculated as a function of $t$ for several different vectors $\mathbf{v}_{\perp}$ which satisfy $\mathbf{v}_{\perp} \cdot \mathbf{v}_{s}=0$ at each instant. The next step is to identify the minimum variance $(\Delta J_{\mathrm{min}})^{2}$ at each instant. From this, 
	the quantity $[1- 2 (\Delta J_{\mathrm{min}})^{2}]$ computed at various instants yields  the degree  of squeezing as a function of time.
	
	Alternatively, we may assess the spin squeezing properties solely from tomograms.  
	In order to compute variances from any generic spin tomogram, we need to express $(\Delta \; \mathbf{J}_{2}\mathbf{\cdot}\mathbf{v}_{\perp})^{2}$ in terms of quantities  that are readily obtained from the relevant tomogram. Using the commutators of the spin operators, it is easy to see that this is possible: 
 $(\mathbf{J}_{2}\mathbf{\cdot}\mathbf{v}_{\perp})^{2}$ involves either terms linear in the 
$\sigma$'s, such as $\sigma_{\smtextsc{a}x}$,  
 or products of $\sigma$'s such as $\sigma_{\smtextsc{a}x}\sigma_{\smtextsc{b}y}$.  
 The expectation values of these operators 
 are readily found. For instance, 
 $\aver{\sigma_{\smtextsc{a}x}}=\sum\limits_{\ma, \mb} \ma w(\mathbf{e_{x}},\ma;\mathbf{e_{y}},\mb)$, while 
  $\aver{\sigma_{\smtextsc{a}x}\sigma_{\smtextsc{b}y}}=\sum\limits_{\ma, \mb} \ma \mb w(\mathbf{e_{x}},\ma;\mathbf{e_{y}},\mb)$.
  
  We adapt the Kitagawa-Ueda squeezing condition in the following manner in order to estimate second-order squeezing (see Sec. 5.3 in~\cite{thesis}). Extending  the argument given above, we consider the expectation value of the dyad $\mathbf{J}_{2}\mathbf{J}_{2}$ instead of $\aver{\mathbf{J}_{2}}$. In general, $\aver{\mathbf{J}_{2}\mathbf{J}_{2}}$ is not a null tensor. We may therefore impose the orthogonality condition  
	$\aver{\mathcal{J}}=0$, where 
	\begin{equation}
	\mathcal{J}=\frac{1}{2}\left(\mathbf{v}_{1}\cdot\mathbf{J}_{2}\mathbf{J}_{2}\cdot \mathbf{v}_{2} + 
	\mathbf{v}_{2}\cdot\mathbf{J}_{2}\mathbf{J}_{2}\cdot \mathbf{v}_{1}\right).
	\label{eqn:2orderSqueezeOp}
	\end{equation}
	$\mathbf{v}_{1}$ and $\mathbf{v}_{2}$ are analogous to the vector $\mathbf{v}_{\perp}$ in the 
	previous  case.  
	The symmetrization with respect to 
	$\mathbf{v}_{1}$ and $\mathbf{v}_{2}$ in 
	Eq.~\eref{eqn:2orderSqueezeOp} ensures that  
	$\mathcal{J}$ is real. We consider a set of several different pairs $(\mathbf{v}_{1},\mathbf{v}_{2})$ for which $\aver{\mathcal{J}(t)}=0$.
	For each such pair, the variance $(\Delta\mathcal{J})^{2}$ has been computed, and from this the minimum variance $(\Delta \mathcal{J}_{\mathrm{min}})^{2}$ has been obtained. 
	The reference value ($= 0.125$) below which the state is second-order squeezed is obtained by minimizing the corresponding variance for the spin coherent state with respect to $\theta$ and $\phi$. 
	
\section{\label{sec:NMRexpt}NMR experiment I}
We now proceed to examine the NMR experiment I.
We will first assess the squeezing properties of the spin system, quantify entanglement with different tomographic entanglement indicators, and comment on the similarities in the dynamics of spin squeezing and entanglement (see Sec. 5.3 in~\cite{thesis}). We also examine the performance of the entanglement indicators in comparison with $\xi_{\smtextsc{qmi}}$ and negativity. As mentioned in Sec.~\ref{sec:intro}, 
%our starting point is the set of reconstructed density matrices at different instants of time obtained from the experiment. T
the system of interest comprises 
$\vphantom{}^{1}\mathrm{H}$ spins (subsystem A),  	
$\vphantom{}^{19}\mathrm{F}$ spins (subsystem B),  	
and 	$\vphantom{}^{13}\mathrm{C}$ spins (subsystem M), evolving in time. 
\begin{figure*}[t]
\centering
\includegraphics[trim=5cm 5cm 5cm 1cm,clip=true,width=12cm]{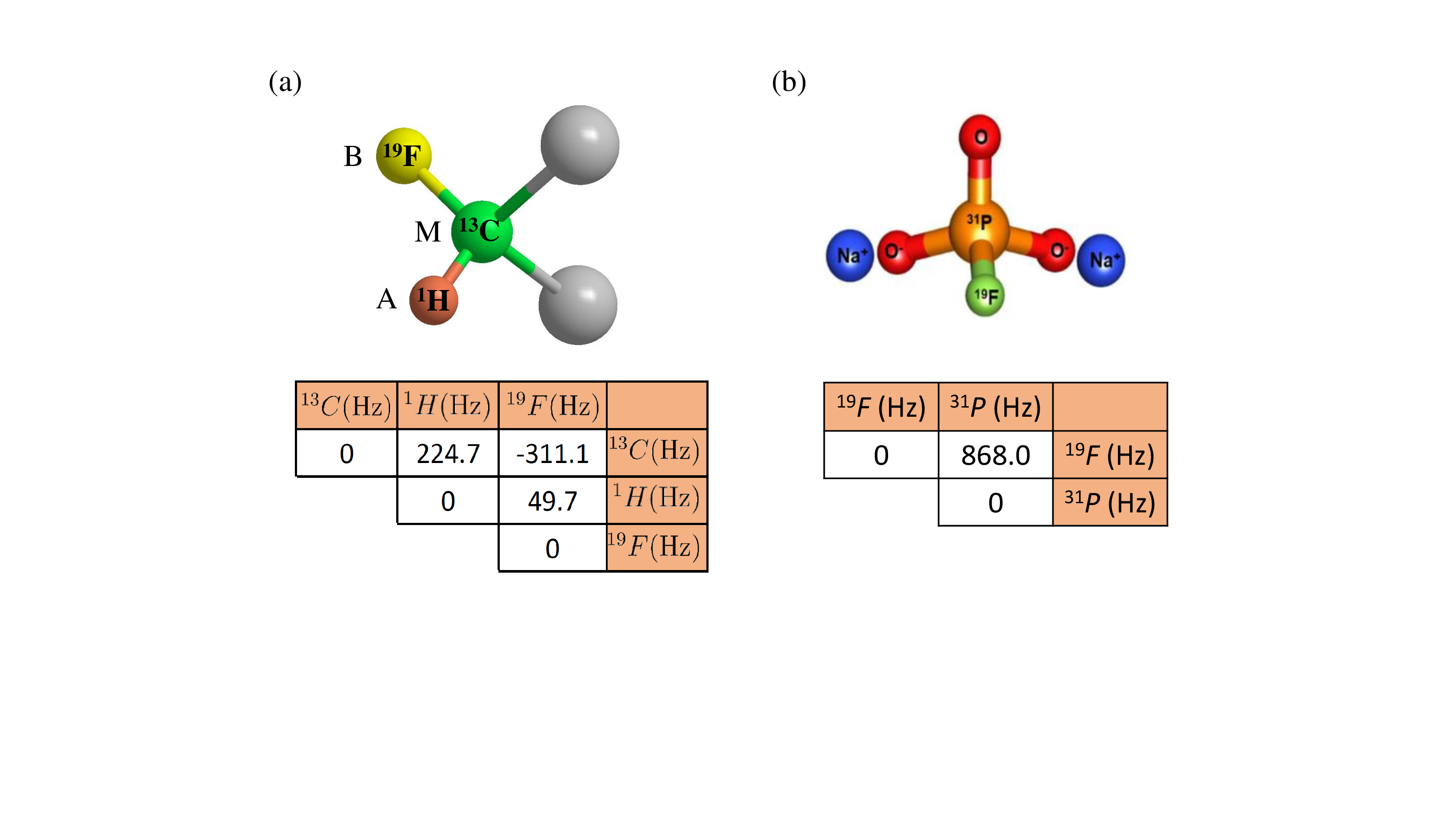}
\caption{Spin systems (a) DBFM molecule and its Hamiltonian parameters forming the three-qubit register used in NMR experiments I and III, (b) NaFP molecule and its Hamiltonian parameters forming the two-qubit register used in NMR experiment II. The off-diagonal elements in each table show the scalar J-coupling values.}
\label{fig:molecule}
\end{figure*}

Experimentally, such a system was realised using the liquid-state Nuclear Magnetic Resonance (NMR) spectroscopy of $^{13}$C, $^{1}$H and $^{19}$F nuclei  in the DBFM molecule.  The liquid sample consisted of an ensemble of around 10$^{15}$ mutually non-interacting  DBFM molecules dissolved in deuterated acetone, placed inside a 11.74 T superconducting magnet. All experiments were performed at ambient temperatures. The nuclear spins interact with the external field via Zeeman interaction which lifts the degeneracy between spin energy levels. Further, inter-molecular interactions are present, but in the liquid state these interactions are all averaged out, and only the intra-molecular interactions mediated via covalent bonds survive \cite{levitt2001spin}. The Hamiltonian of the system in a frame rotating about the quantization axis with respective Larmor frequencies can be written (in units of $\hbar$) as
\begin{equation}
H_{\smtextsc{i}}=   
\frac{1}{2}\pi  \big ( J_{\smtextsc{am}} \, \sigma_{\smtextsc{a} z} \otimes \sigma_{\smtextsc{m} z} +  J_{\smtextsc{bm}} \, \sigma_{\smtextsc{b} z} \otimes \sigma_{\smtextsc{m} z} +
J_{\smtextsc{ab}} \, \sigma_{\smtextsc{a} z} \otimes \sigma_{\smtextsc{b} z} \big ),~~~~~~
\label{eqn:HJ1}
\end{equation}
where the coupling constants 
$\{J\}$  are as indicated in Fig.~\ref{fig:molecule} (a). %$J_{\smtextsc{am}}=224.5 ~\mathrm{Hz}~,J_{\smtextsc{bm}}=-310.9 ~\mathrm{Hz},~\mathrm{and}~J_{\smtextsc{ab}}=49.7 ~\mathrm{Hz}$ are the corresponding scalar coupling constants (capturing the intra-molecular interactions) between the nuclei. %We set all $\Omega_i = 0$ by moving to a rotating frame at the individual Larmor frequencies.
The interaction between the nuclei, along with  radio-frequency (RF) pulses can be  used to realise the effective Hamiltonian 
\begin{equation}
H_{\smtextsc{s}}=4 \chi_{\mathrm{s}} (\sigma_{\smtextsc{a} x} + \sigma_{\smtextsc{b} x}) \sigma_{\smtextsc{m} x},
\label{eqn:Ham_spin}
\end{equation}
where $\chi_{\mathrm{s}}$ is a constant~ 
\cite{nmrExpt}. Thus, 
in each DBFM molecule, the focus is on a chain of three qubits, in the topology A-M-B, such that the probe qubits  A and B interact with each other only via the 
mediator qubit M.
 The three qubit system is then prepared in the initial state 
 given by the density matrix
\begin{equation}
\rho_{\smtextsc{mab}}(0)=
\frac{1}{2} \ket{\phi_{+}}_{\smtextsc{ab ab}} \hspace*{-0.25 em}\bra{\phi_{+}} \otimes \rho_{\smtextsc{m} +} + 
\frac{1}{2} \ket{\psi_{+}}_{\smtextsc{ab ab}} \hspace*{-0.25 em}\bra{\psi_{+}} \otimes \rho_{\smtextsc{m} -},
\label{eqn:rhozero_defn}
\end{equation}
where $\rho_{\smtextsc{m} +}=\ket{+}_{\smtextsc{m m}} \hspace*{-0.25 em}\bra{+}$, $\rho_{\smtextsc{m} -}=\ket{-}_{\smtextsc{m m}} \hspace*{-0.25 em}\bra{-}$, and
\begin{equation}
\left.
\begin{array}{lll}
\ket{\psi_{+}}_{\smtextsc{ab}}&=&\big(\ket{\downarrow}_{\smtextsc{a}}\otimes\ket{\downarrow}_{\smtextsc{b}} + \ket{\uparrow}_{\smtextsc{a}}\otimes\ket{\uparrow}_{\smtextsc{b}}\big)/\sqrt{2}, \\[4pt]
\ket{\phi_{+}}_{\smtextsc{ab}}&=&
\big(\ket{\downarrow}_{\smtextsc{a}}\otimes\ket{\uparrow}_{\smtextsc{b}} + \ket{\uparrow}_{\smtextsc{a}}\otimes\ket{\downarrow}_{\smtextsc{b}}\big)/\sqrt{2}.
\end{array}
\right\}
\label{eqn:phi0_defn}
\end{equation}
Here, $\ket{\uparrow}$ and $\ket{\downarrow}$ denote
 the eigenstates of $\sigma_{z}$, while 
$\ket{+}$ and $\ket{-}$ denote those of $\sigma_{x}$. 
It is straightforward to show that
\begin{equation}
\rho_{\smtextsc{mab}}(t) = 
\frac{1}{2} \ket{\Psi_{0}}\bra{\Psi_{0}} + \frac{1}{2} \ket{\Psi_{1}}\bra{\Psi_{1}}
\label{eqn:rhot_defn} 
\end{equation}
where
\begin{equation}
\left.
\begin{array}{lll}
\ket{\Psi_{0}}&=&\ket{+}_{\smtextsc{m}} \left[\cos (2 \chi_{\mathrm{s}} t) \ket{\phi_{+}}_{\smtextsc{ab}} - i \sin (2 \chi_{\mathrm{s}} t) \ket{\psi_{+}}_{\smtextsc{ab}}\right], \\[4pt] 
\ket{\Psi_{1}}&=&\ket{-}_{\smtextsc{m}} \left[\cos (2 \chi_{\mathrm{s}} t) \ket{\psi_{+}}_{\smtextsc{ab}} + i \sin (2 \chi_{\mathrm{s}} t) \ket{\phi_{+}}_{\smtextsc{ab}}\right].
\end{array}
\right\}	
\label{eqn:Psi1}
\end{equation}
Here, and in the other two experiments reported in the subsequent section, we have carried out the following procedure. 
(a) We have used the experimentally-obtained tomograms, and computed numerical values of the extents of entanglement and squeezing directly, at different instants of time. 
(b) At these instants, we have numerically obtained the density matrices using the Liouville equation, with the initial state and the Hamiltonian as inputs, and recreated the corresponding tomograms. From the latter, we have computed the extent of entanglement and squeezing. This approach gives us an estimate of the experimental losses. (c) Further, direct computation of the standard entanglement measures ($\xi_{\textsc{svne}}$, $\xi_{\textsc{qmi}}$) from both the reconstructed and numerically-obtained density matrices have been carried out. This facilitates assessment of the efficacy of calculations performed with tomograms without resorting to density matrices. Since the focus in this paper is to assess the advantages of the tomographic approach vis-a-vis the density matrix approach, we have used the experimentally-obtained tomogram discounting the error bars. This suffices for our purpose.

Of direct relevance to us is the reduced  density matrix 
\begin{equation}
\rho_{\smtextsc{ab}}(t)=\mathrm{Tr}_{\smtextsc{m}}(\rho_{\smtextsc{mab}}(t))
\label{eqn:RhoABnmr}
\end{equation} 
corresponding to the subsystem AB.

\begin{figure}[t]
\centering
\includegraphics[width=0.32\textwidth]{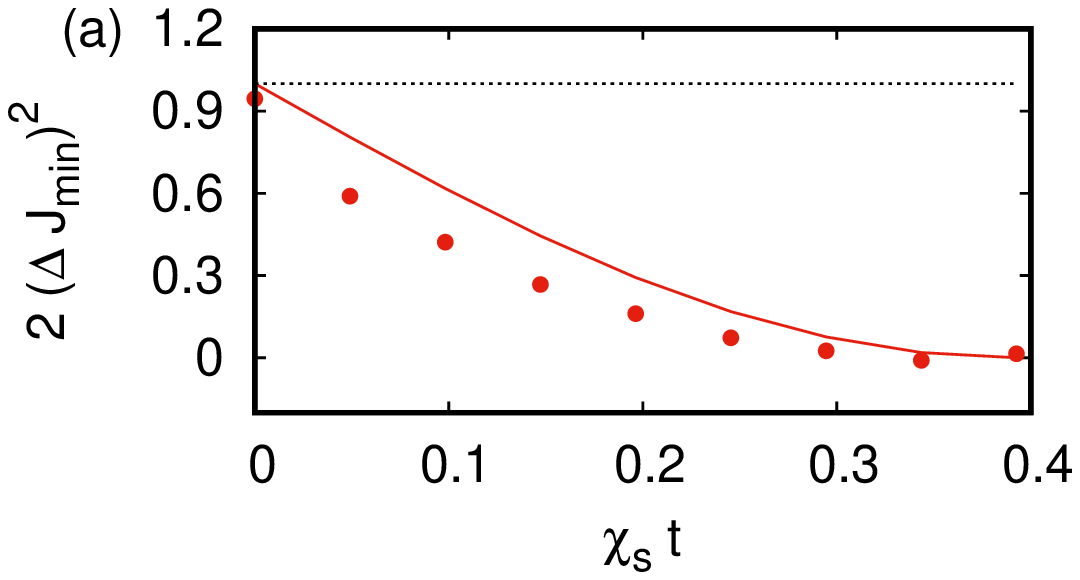}
\includegraphics[width=0.32\textwidth]{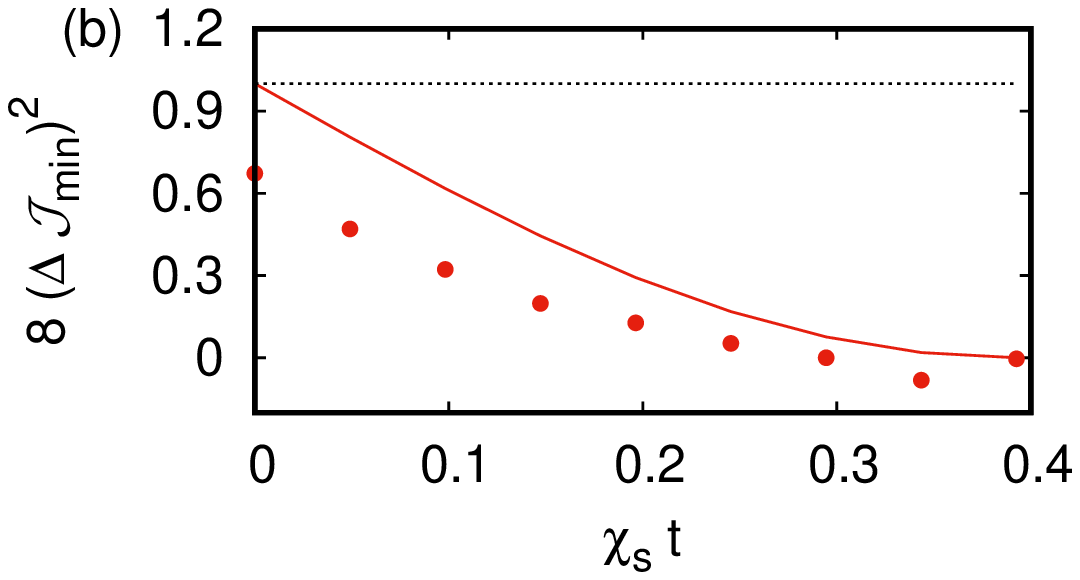}
\caption{(a) $2 (\Delta J_{\mathrm{min}})^{2}$\, (b) $8 (\Delta \mathcal{J}_{\mathrm{min}})^{2}$ vs. scaled time $\chi_{\mathrm{s}} t$. The solid curves are computed using Eq.~\eref{eqn:RhoABnmr} and the dotted curves from experimental data. The black horizontal line  in (a) and  (b) sets the limit below which the state is squeezed.}
\label{fig:squeeze_compare}
\end{figure}

We now proceed to examine the spin squeezing properties, and as a first step we compute the mean spin direction $\mathbf{v}_{\smtextsc{s}}(t)$ in the NMR experiment I. Since $\aver{\sigma_{i x}(t)}$, $\aver{\sigma_{i y}(t)}$ and $\aver{\sigma_{i z}(t)}$ ($i=\mathrm{A,B}$) are all
equal to zero, it follows that $\mathbf{v}_{\smtextsc{s}}(t)$ is a null vector. Hence, any unit vector $\mathbf{v}_{\perp}$ can be chosen to obtain the required variance. We have calculated 
the variance $(\Delta \; \mathbf{J}_{2}\mathbf{\cdot}\mathbf{v}_{\perp})^{2} = \aver{(\mathbf{J}_{2}\mathbf{\cdot}\mathbf{v}_{\perp})^{2}}$ as a function of $t$  for $800$ different vectors $\mathbf{v}_{\perp}$ at each instant. From this, we have identified the minimum variance $(\Delta J_{\mathrm{min}})^{2}$ and plotted it as a function of time (Fig.~\ref{fig:squeeze_compare}(a)). From the figure, it is evident that the variance obtained numerically using Eq.~\eref{eqn:RhoABnmr} and 
that obtained from the experimentally reconstructed density matrices are in good agreement. We also point out that the extent of squeezing,  $[1-2 (\Delta J_{\mathrm{min}})^{2}]$,  increases with time. We have also verified that the variances obtained from the tomograms agree with those computed from the corresponding density matrix.
	
In order to estimate second-order squeezing using the procedure given in Sec.~\ref{sec:squeeze_prop}, we first compute $\aver{\mathcal{J} (t)}$. Using Eq.~\eref{eqn:RhoABnmr}, we get 
\begin{align}
\nonumber\aver{\mathcal{J} (t)}&=\mathrm{Tr}\,\big(\rho_{\smtextsc{ab}}(t) \mathcal{J}\big)\\
\nonumber & = \mathbf{v}_{1 x} \mathbf{v}_{2 x} +
\frac{1}{2} \big\{ \mathbf{v}_{1y} \mathbf{v}_{2 y} + \mathbf{v}_{1z} \mathbf{v}_{2 z} \\
&+ \sin (4 \chi_{\mathrm{s}} t) \big( \mathbf{v}_{1 y} \mathbf{v}_{2 z} + \mathbf{v}_{2 y} \mathbf{v}_{1 z}\big)\big\}, 
\label{eqn:avercalJ}
\end{align}
where the subscripts $x$, $y$ and $z$ denote the corresponding  components of $\mathbf{v_{1}}$ and $\mathbf{v_{2}}$. (The symmetry between the $y$ and $z$ components in Eq.~\eref{eqn:avercalJ} follows from the fact that $[\sigma_{\smtextsc{a} x}\sigma_{\smtextsc{b} x},\rho_{\smtextsc{ab}}(t)]=0$.) We have considered a set of $320$ different pairs $(\mathbf{v}_{1},\mathbf{v}_{2})$ for which $\aver{\mathcal{J}(t)}=0$.
For each such pair, the variance $(\Delta\mathcal{J})^{2}$ has been computed, and from this the minimum variance $(\Delta \mathcal{J}_{\mathrm{min}})^{2}$ has been obtained. 
Plots of $(\Delta \mathcal{J}_{\mathrm{min}})^{2}$ versus $t$  obtained both from the experimentally reconstructed density matrices and from Eq.~\eref{eqn:RhoABnmr} are shown in Fig.~\ref{fig:squeeze_compare}(b). 
The two  curves are in reasonable agreement with each other. As in the earlier case, the measure of second-order squeezing, $[1-8 (\Delta \mathcal{J}_{\mathrm{min}})^{2}]$,  increases with time.
We have verified that neither the state of subsystem A nor that of B displays entropic squeezing~\cite{MassenEUR} at any time.
	
We turn now to the entanglement dynamics in NMR experiment I. We have computed $\xi_{\smtextsc{tei}}$, $\xi_{\smtextsc{ipr}}$, $\xi_{\smtextsc{bd}}$ and $\xi_{\smtextsc{pcc}}$, and compared with two standard indicators, $\xi_{\smtextsc{qmi}}$ and the negativity $N(\rho_{\smtextsc{ab}})$. (Negativity has been computed from density matrices in the experiment~\cite{nmrExpt}, without using the tomographic approach). Negativity is defined as  $N(\rho_{\smtextsc{ab}})=
\frac{1}{2}\sum_{i} \left(\vert\mathcal{L}_{i}\vert-\mathcal{L}_{i}\right)$. 
Here $\lbrace\mathcal{L}_{i}\rbrace$ is the set 
of  eigenvalues of $\rho^{T_{\smtextsc{a}}}
_{\smtextsc{ab}}$, 
the partial transpose of $\rho_{\smtextsc{ab}}$ with respect to the subsystem A. (Equivalently, 
the  partial transpose  
$\rho^{T_{\smtextsc{b}}}_{\smtextsc{ab}}$  
may be used.)  
\begin{figure*}[t]
\centering
\includegraphics[width=0.4\textwidth]{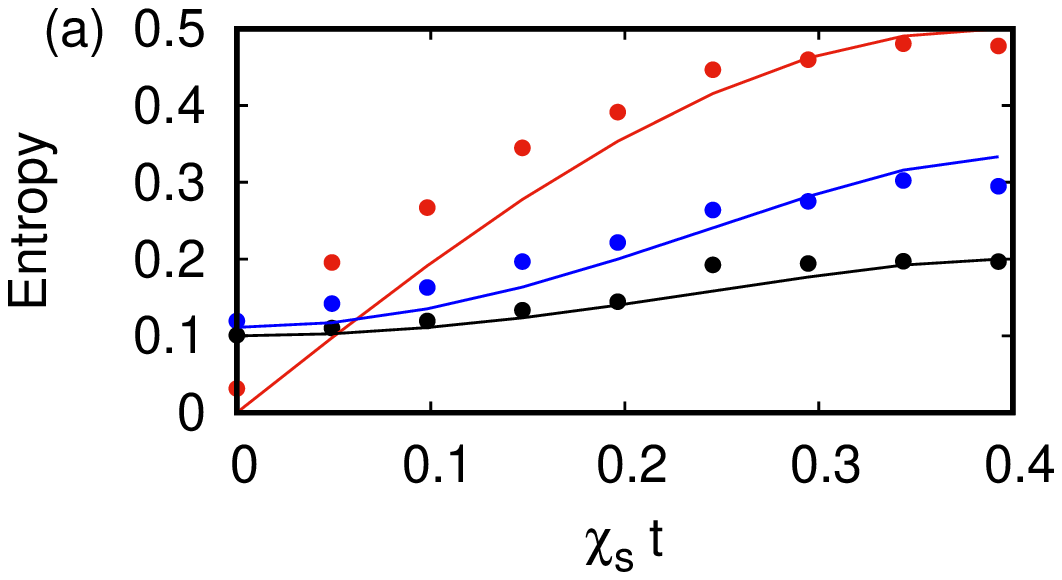}
\includegraphics[width=0.4\textwidth]{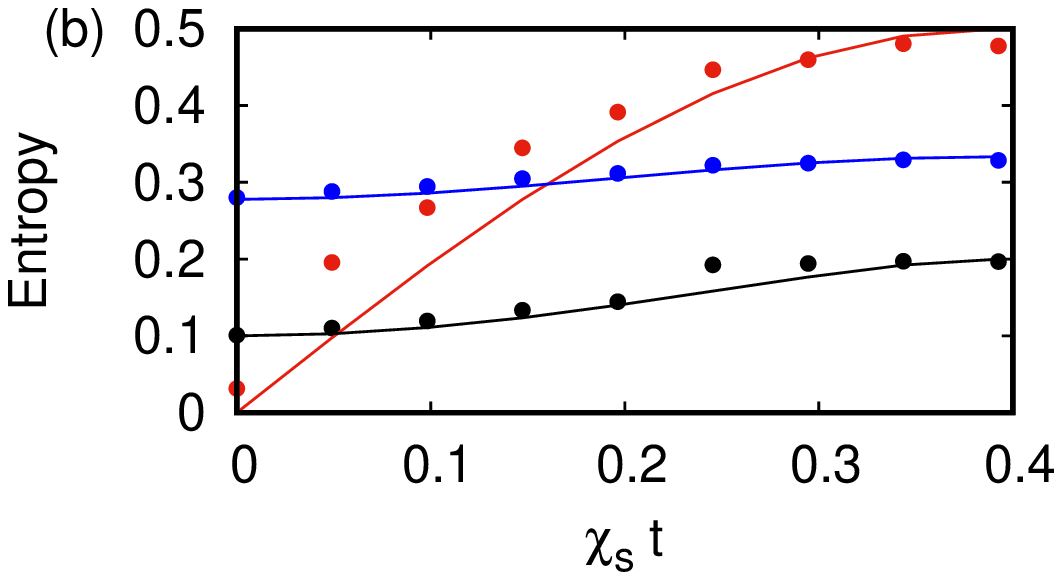}
\includegraphics[width=0.4\textwidth]{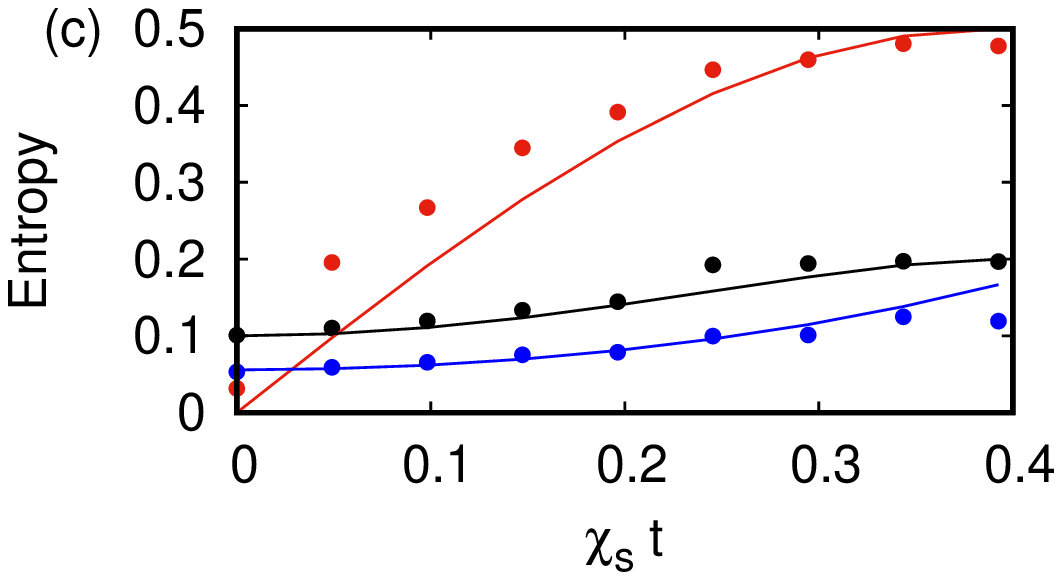}
\includegraphics[width=0.4\textwidth]{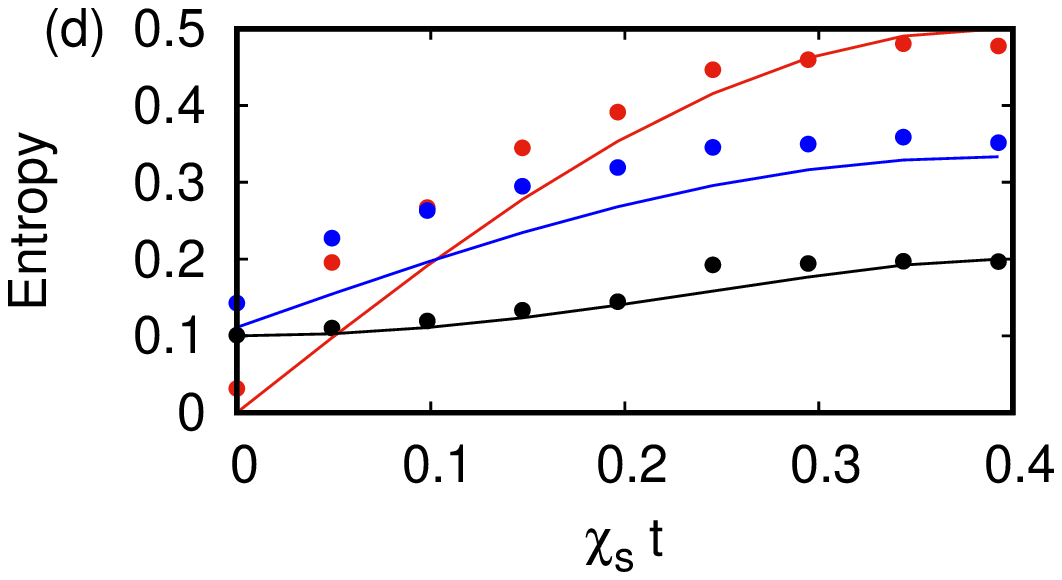}
\caption{$N(\rho_{\smtextsc{ab}})$ (red), $0.1 \,\xi_{\smtextsc{qmi}}$ (black), and (a) $\xi_{\smtextsc{tei}}$ (blue), (b) $\xi_{\smtextsc{ipr}}$ (blue), (c) $\xi_{\smtextsc{bd}}$ (blue), (d) $\xi_{\smtextsc{pcc}}$ (blue) vs. scaled time $\chi_{\mathrm{s}} t$. The solid curves are computed using Eq.~\eref{eqn:RhoABnmr} and the dotted curves from experimental data.}
\label{fig:indics_negativity_compare}
\end{figure*}
We find that $\xi_{\smtextsc{tei}}$, $\xi_{\smtextsc{qmi}}$ and $N(\rho_{\smtextsc{ab}})$
are in agreement  in their gross features (Fig.~\ref{fig:indics_negativity_compare} (a)), with $\xi_{\smtextsc{tei}}$ closer to $\xi_{\smtextsc{qmi}}$ 
due to the similarity in their definitions.
The indicators $\xi_{\smtextsc{bd}}$ and $\xi_{\smtextsc{ipr}}$ behave in a manner similar to $\xi_{\smtextsc{tei}}$ (Figs.~\ref{fig:indics_negativity_compare} (b) and (c)). Surprisingly, the indicator $\xi_{\smtextsc{pcc}}$ that captures only linear correlations, performs well, and agrees reasonably well with $N(\rho_{\smtextsc{ab}})$ (Fig.~\ref{fig:indics_negativity_compare} (d)). This is in contrast to earlier results on bipartite CV and multipartite HQ systems~\cite{sharmila4}, where it was shown that $\xi_{\smtextsc{pcc}}$ does not reliably track  the extent of entanglement. 
%%%%%%%%%%%%%%%%%%%%%

\begin{figure}
\centering
\includegraphics[width=0.4\textwidth]{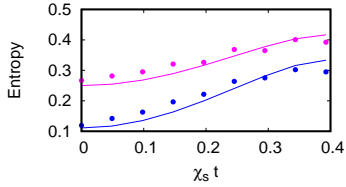}
\caption{$\xi_{\smtextsc{tei}}$ computed from 9 tomographic slices (blue) and $\xi^{\prime}_{\smtextsc{tei}}$ computed from 6 tomographic slices (magenta) vs. scaled time $\chi_{\mathrm{s}} t$. The solid curves are computed using Eq.~\eref{eqn:RhoABnmr} and the dotted curves from experimental data.}
\label{fig:TEI_compare}
\end{figure}

Typically, our computations involve tomograms obtained over $9$ basis sets, namely, $x x$, $x y$, $x z$, $y x$, $y y$, $y z$, $z x$, $z y$, and $z z$. To estimate the extent of squeezing, we require all 9 `tomographic slices' corresponding to these $9$ basis sets, as is evident from our procedure to compute $2 (\Delta J_{\mathrm{min}})^{2}$ and $8 (\Delta \mathcal{J}_{\mathrm{min}})^{2}$ in Sec.~\ref{sec:squeeze_prop}. However, a natural question that arises is whether the behaviour of the entanglement indicators can be captured from a smaller  number of tomographic slices. In NMR experiment I in Fig.~\ref{fig:TEI_compare}, we compare $\xi_{\smtextsc{tei}}$ obtained using the original 9 slices with $\xi^{\prime}_{\smtextsc{tei}}$ which is a similar averaged indicator computed with only 6 slices (namely,  $x x$, $x y$, $x z$, $y x$, $y y$, and $y z$). With this choice of the reduced number of slices, we see from Fig.~\ref{fig:TEI_compare} that $\xi_{\smtextsc{tei}}$ and $\xi^{\prime}_{\smtextsc{tei}}$ are in good agreement with each other. We have verified that this does not hold for other indicators such as $\xi_{\smtextsc{ipr}}$. A naive extrapolation of our results would imply that for the bipartite system in NMR experiment II, too,  a smaller  number of tomographic slices would suffice. However, as shown in the following section, that is not possible.

\begin{figure}[t]
\centering
\includegraphics[width=0.4\textwidth]{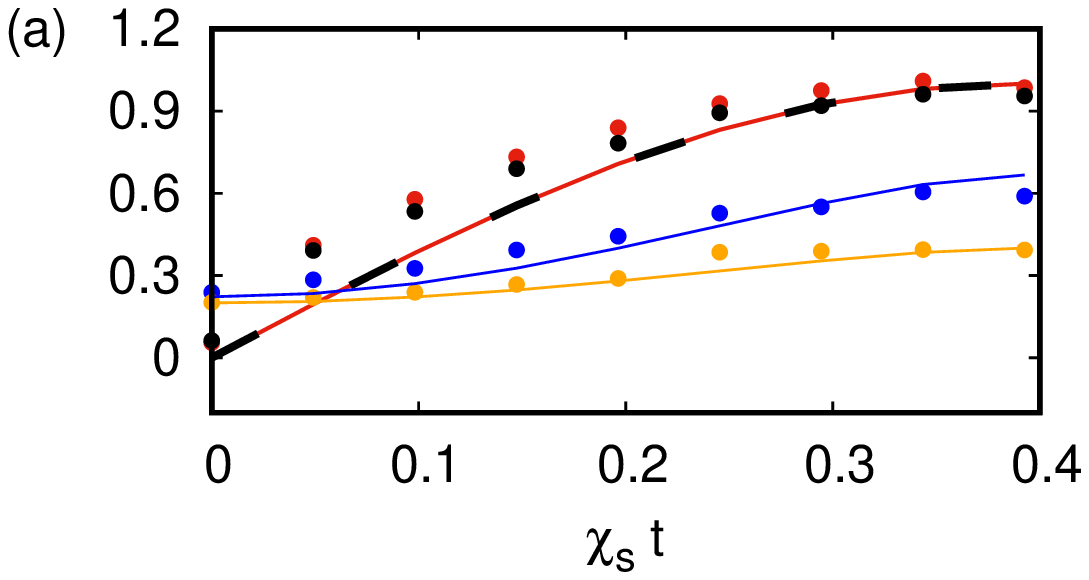}
\includegraphics[width=0.4\textwidth]{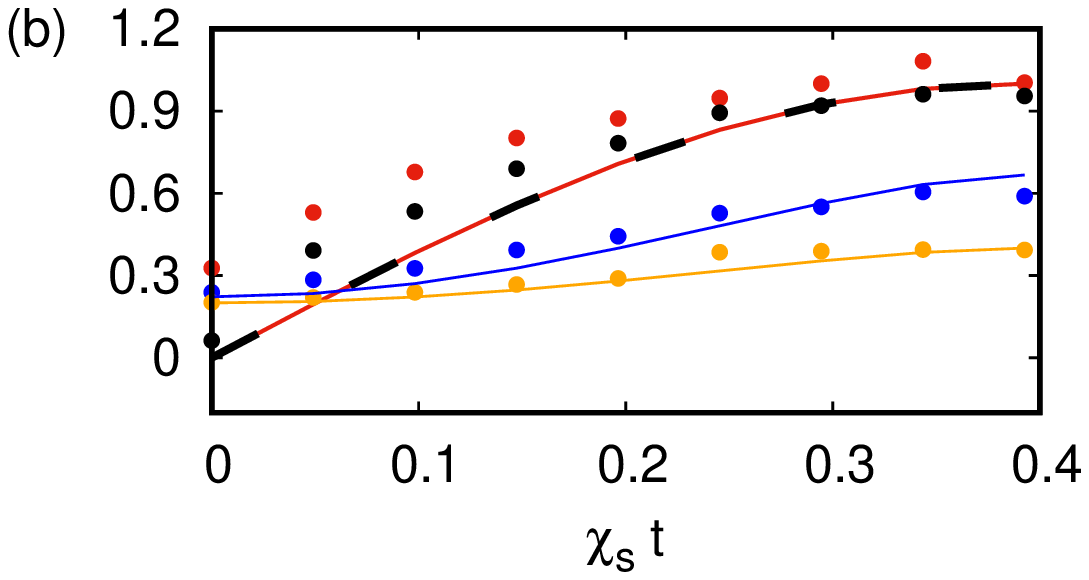}
\caption{$2 \, N(\rho_{\smtextsc{ab}})$ (black), $2 \, \xi_{\smtextsc{tei}}$ (blue), $0.2 \,\xi_{\smtextsc{qmi}}$ (orange), and (a) $[1 - 2 (\Delta J_{\mathrm{min}})^{2}]$ (red), (b) $[1 - 8 (\Delta \mathcal{J}_{\mathrm{min}})^{2}]$ (red), vs. scaled time $\chi_{\mathrm{s}} t$. The solid curves are computed using Eq.~\eref{eqn:RhoABnmr} and the dotted curves from experimental data.}
\label{fig:squeeze_extent_tei_compare}
\end{figure}

\begin{figure}[t]
\centering
\includegraphics[width=0.4\textwidth]{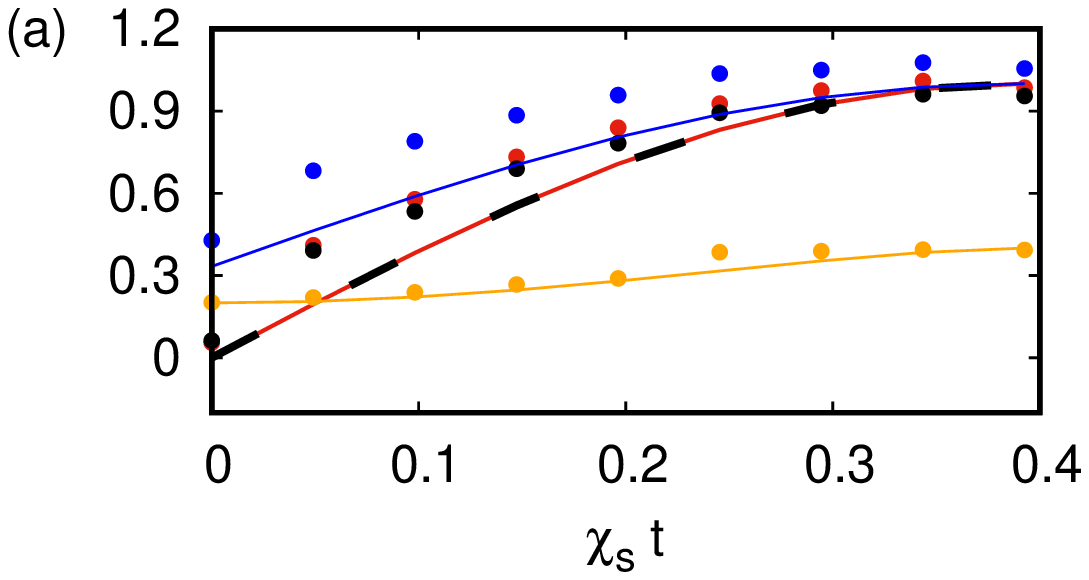}
\includegraphics[width=0.4\textwidth]{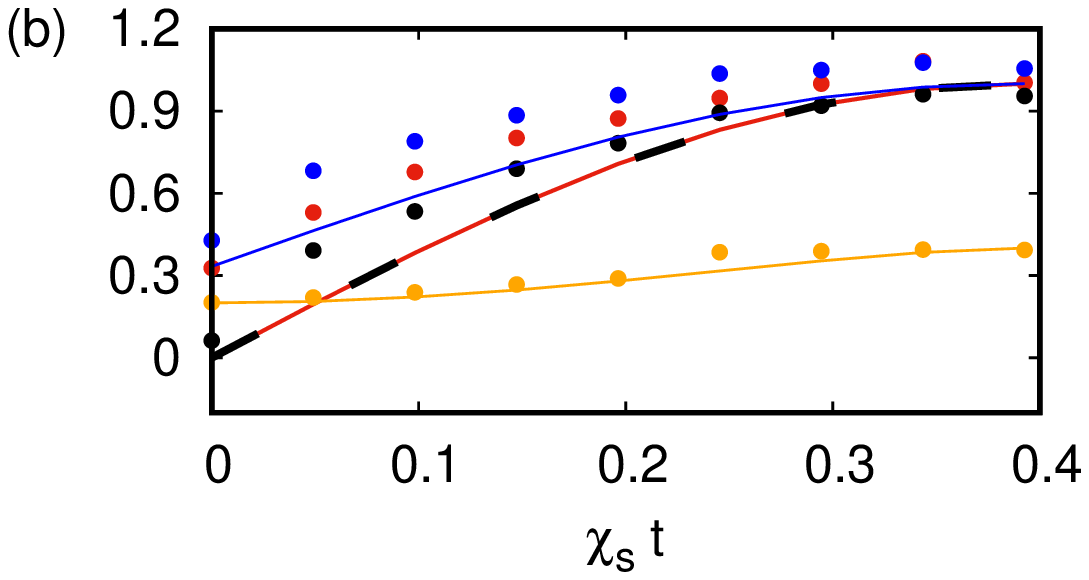}
\caption{$2 \, N(\rho_{\smtextsc{ab}})$ (black), $3 \, \xi_{\smtextsc{pcc}}$ (blue), $0.2 \,\xi_{\smtextsc{qmi}}$ (orange), and (a) $[1 - 2 (\Delta J_{\mathrm{min}})^{2}]$ (red), (b) $[1 - 8 (\Delta \mathcal{J}_{\mathrm{min}})^{2}]$ (red), vs. scaled time $\chi_{\mathrm{s}} t$. The solid curves are computed using Eq.~\eref{eqn:RhoABnmr} and the dotted curves from experimental data.}
\label{fig:squeeze_extent_pcc_compare}
\end{figure}

\begin{figure}[t]
\centering
\includegraphics[width=0.4\textwidth]{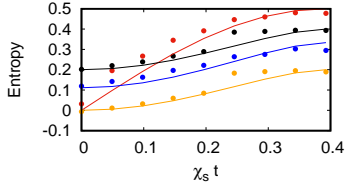}
\caption{$N(\rho_{\smtextsc{ab}})$ (red), $0.2 \,\xi_{\smtextsc{qmi}}$ (black), $\xi_{\smtextsc{tei}}$ (blue),  and $0.2\,D(\mathrm{A}:\mathrm{B})$ (orange) vs. scaled time $\chi_{\mathrm{s}} t$. The solid curves are computed using Eq.~\eref{eqn:RhoABnmr} and the dotted curves from experimental data.}
\label{fig:discord_compare}
\end{figure}

Returning to the problem at hand, Figs. \ref{fig:squeeze_extent_tei_compare} (a) and (b) facilitate comparison between $[1-2 (\Delta J_{\mathrm{min}})^{2}]$, $[1-8 (\Delta \mathcal{J}_{\mathrm{min}})^{2}]$, $N(\rho_{\smtextsc{ab}})$, $\xi_{\smtextsc{tei}}$, and $\xi_{\smtextsc{qmi}}$.
It is clear that $N(\rho_{\smtextsc{ab}})$ characterises the degree of squeezing and higher-order squeezing extremely well. $\xi_{\smtextsc{tei}}$ and $\xi_{\smtextsc{qmi}}$ are approximate estimators of squeezing properties. However, we see from Figs.~\ref{fig:squeeze_extent_pcc_compare} (a) and (b) that  $\xi_{\smtextsc{pcc}}$ compares well with $[1-2 (\Delta J_{\mathrm{min}})^{2}]$, $[1-8 (\Delta \mathcal{J}_{\mathrm{min}})^{2}]$, and $N(\rho_{\smtextsc{ab}})$. The variances and covariances seem to  capture the behaviour of $N(\rho_{\smtextsc{ab}})$ quite well,
while $\xi_{\smtextsc{tei}}$ reflects the behaviour of $\xi_{\smtextsc{qmi}}$. For completeness, we also present Fig.~\ref{fig:discord_compare} comparing $\xi_{\smtextsc{tei}}$, $\xi_{\smtextsc{qmi}}$, and $N(\rho_{\smtextsc{ab}})$ with discord $D(\mathrm{A}:\mathrm{B})$. We see that $\xi_{\smtextsc{tei}}$ agrees well with both $\xi_{\smtextsc{qmi}}$, and $D(\mathrm{A}:\mathrm{B})$. Our calculations also reveal that $\xi_{\smtextsc{pcc}}$ is not in good agreement with discord. Equivalent circuits corresponding to subsystem AB of this NMR experiment at various instants of time have been processed through the IBM Q platform \cite{ibm_main}. The entanglement indicators have been obtained through this exercise. The details are reported in the Appendix. 

%%%%%%%%%IBMQ%%%%%%%%%%%%%%MOVE TO APPENDIX%%
	
%%%%%%%%%%%%%%%%%%%%%%%%%%%%%%%%%%%%%%%%%%%%%%%%%%%%%%%%%%%%%%%%%%%%
\section{\label{sec:NMRexpt2}Blockade and freezing in nuclear spins}

We now proceed to investigate NMR experiments II and III. As mentioned in Sec.~\ref{sec:intro}, the systems of interest in this case, comprise $N$ spin qubits ($N=2$ for II, $N=3$ for III), and are examined using NMR spectroscopic techniques as they evolve in time. In both cases, the extent of bipartite entanglement has been reported, using discord as the entanglement indicator. In NMR experiment II, each qubit is a subsystem, and in experiment III, one subsystem comprises two qubits and the other has a single qubit. The effective Hamiltonian for $N$ qubits is
\begin{equation}
H_{N}= \sum_{i=1}^{N} \left( \omega_{i} \sigma_{i x} - \Omega_{i} \sigma_{i z}\right) + \sum_{i=1}^{N-1} \sum_{j=i+1}^{N} \lambda_{ij} \sigma_{i z} \sigma_{j z},
\label{eqn:Ham_nmr2}
\end{equation}
where $\omega_{i}$, $\Omega_{i}$ and $\lambda_{ij}$ are constants, $\sigma_{x}$ and $\sigma_{z}$ are the usual spin matrices, and the subscripts $i, j$ label the corresponding qubits.
	
The density matrix at time $t=0$ is 
\begin{equation}
\rho_{N}(0)=\Big(\frac{1-\epsilon}{2^{N}}
\Big)\mathbb{I}_{N} + \epsilon \ket{\psi} \bra{\psi},
\label{eqn:rhoNinit}
\end{equation}
where $\mathbb{I}_{N}$ is the identity operator of dimension $2^{N} \times 2^{N}$, $\ket{\psi}=\ket{\downarrow}^{\otimes N}$, and $\epsilon$ is the purity of the state. Therefore, in principle,  $\text{Tr}[\rho_{N}^2]$ lies between 
the values $1$ (when $\epsilon=1$) and $1/2^N$ (when $\epsilon=0$). The time-evolved state is obtained numerically. We compute $\xi_{\smtextsc{tei}}$, $\xi_{\smtextsc{ipr}}$ and $\xi_{\smtextsc{bd}}$ for both the systems from the corresponding tomograms and examine the extent to which they capture the entanglement features seen in the discord as a function of time. We also comment on the spin squeezing properties.

\subsection{\label{subsec:bipart}NMR experiment II}

As mentioned in Sec.~\ref{sec:intro}, the experiment has been performed on two nuclear spins using 
$\vphantom{}^{19}\mathrm{F}$ and $\vphantom{}^{31}\mathrm{P}$ spins in NaFP (Fig.~\ref{fig:molecule} (b)). Here $\vphantom{}^{19}\mathrm{F}$  and 
$\vphantom{}^{31}\mathrm{P}$  are subsystems 
1 and 2 respectively.  
The internal Hamiltonian for the two qubit system in a doubly rotating frame in the absence of any RF drive is given by
\begin{equation}
H_{\smtextsc{ii}} = -\sum\limits_{i=1}^{2}\Omega_i \sigma_{i z} + \frac{1}{2}\pi  \big ( J_{\smtextsc{fp}} \,\sigma_{1 z} \otimes \sigma_{2 z}\big ),
\label{Hnmr2q}
\end{equation}
where $J_{\smtextsc{fp}}$ is the scalar coupling constant. In order to realize blockade and freezing phenomena, the system is further driven by RF pulses along the $x$-axis. The effective Hamiltonian is then given by
\begin{equation}
H_{\smtextsc{ii}}^{\prime}= \sum\limits_{i=1}^{2} \omega_{i} \sigma_{i x} + H_{\smtextsc{ii}},
\label{eqn:Hnmr2qRyd}
\end{equation}
where $\omega_i$ are the corresponding drive amplitudes. The constants $\omega_{i}, \Omega_{i}$ and $\pi J_{\smtextsc{fp}}/2$ in Eq.~\eref{eqn:Hnmr2qRyd} can be directly mapped to the constants  $\omega_{i}, \Omega_{i}$ and $\lambda_{ij}$, respectively, in Eq.~\eref{eqn:Ham_nmr2}. The system is prepared in a pseudo-pure state, as defined in Eq.~\eref{eqn:rhoNinit} with $N=2$, which is isomorphic to the pure state $\ket{\downarrow}^{\otimes 2}$. Corresponding to this initial state, the discord has been computed for $N=2$ 
as a function of time, and reported for 
$\lambda_{12}/(2 \pi) = 868$ Hz, and $\Omega_{1}=\Omega_{2}=\lambda_{12}/2$. Three cases have been examined, namely, (i) $\omega_{1}/(2 \pi)=217$ Hz, $\omega_{2}=\omega_{1}$ (which describes the blockade condition), 
and (ii) $\omega_{2}/(2 \pi)=217$ Hz, $\omega_{1}=\omega_{2}/4$,  (iii) $\omega_{1}/(2 \pi)=217$ Hz, $\omega_{2}=\omega_{1}/4$, (which describe freezing as detailed in \cite{nmrExpt1}). Following a similar procedure, we have now computed $\xi_{\smtextsc{qmi}}$ 
as a function of time. This indicator has been obtained both from numerical computations and from the experimentally reconstructed density matrices. In order to examine the performance of the tomographic indicators, we have used the tomograms corresponding to both these density matrices at each instant of time. From these, the corresponding $\xi_{\smtextsc{tei}}$, $\xi_{\smtextsc{ipr}}$, and $\xi_{\smtextsc{bd}}$ are computed. 
	
\begin{figure*}[t]
\xincludegraphics[width=0.45\textwidth,label=(a),fontsize=\scriptsize]{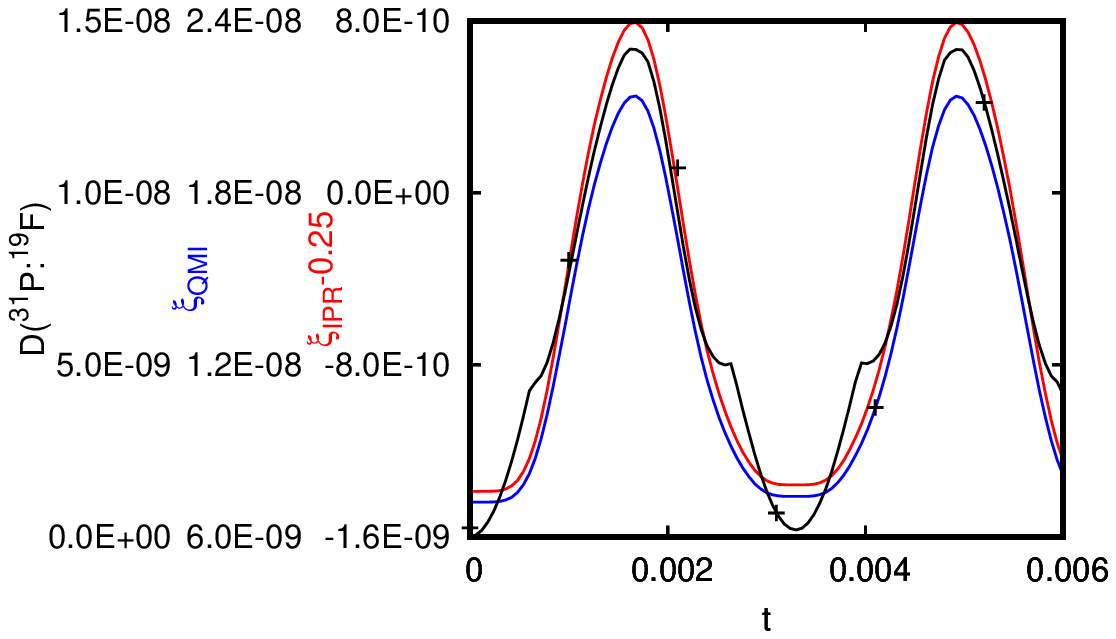}
\xincludegraphics[width=0.45\textwidth,label=(b),fontsize=\scriptsize]{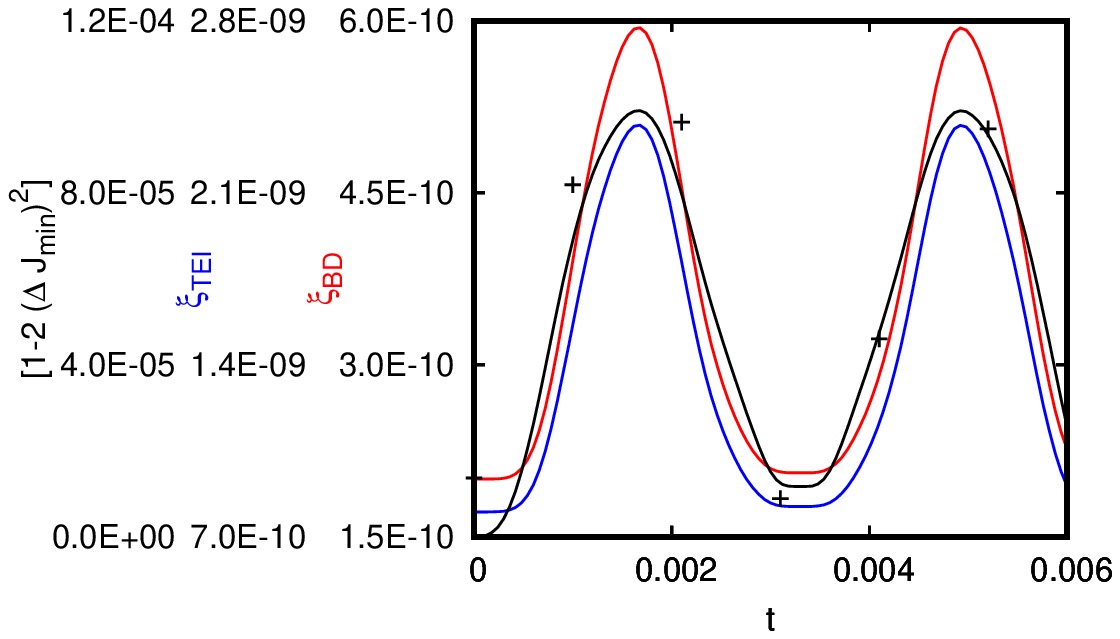}
%\xincludegraphics[width=0.31\textwidth,label=(c),fontsize=\scriptsize]{bipartite_blockade_BD.eps}
\caption{
(a) $D( {}^{31}\mathrm{P} : {}^{19}\mathrm{F})$ (black), $\xi_{\smtextsc{qmi}}$ (blue), and $\xi_{\smtextsc{ipr}}$ (red) vs. time $t$ in seconds, and (b) $[1 - 2 (\Delta J_{\mathrm{min}})^{2}]$ (black), $\xi_{\smtextsc{tei}}$ (blue), and $\xi_{\smtextsc{bd}}$ (red) vs. time $t$ in seconds for Case (i). The solid curves are computed by numerical simulation and the crosses from experimental data.
}
\label{fig:bipart_block}
\end{figure*}

\begin{figure*}[t]
\xincludegraphics[width=0.45\textwidth,label=(a),fontsize=\scriptsize]{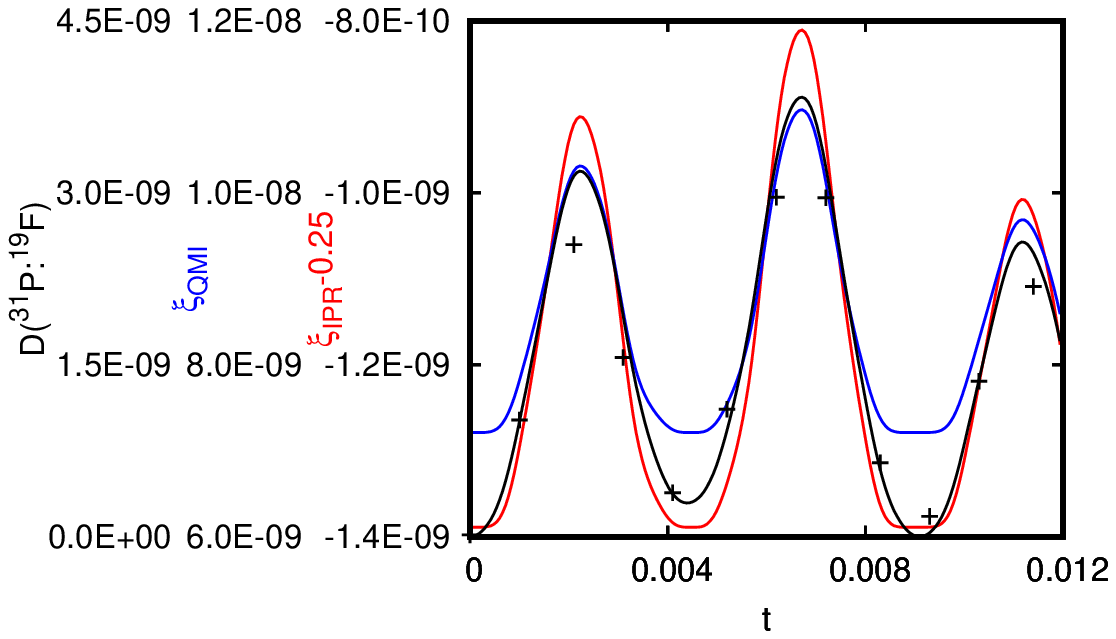}
\xincludegraphics[width=0.45\textwidth,label=(b),fontsize=\scriptsize]{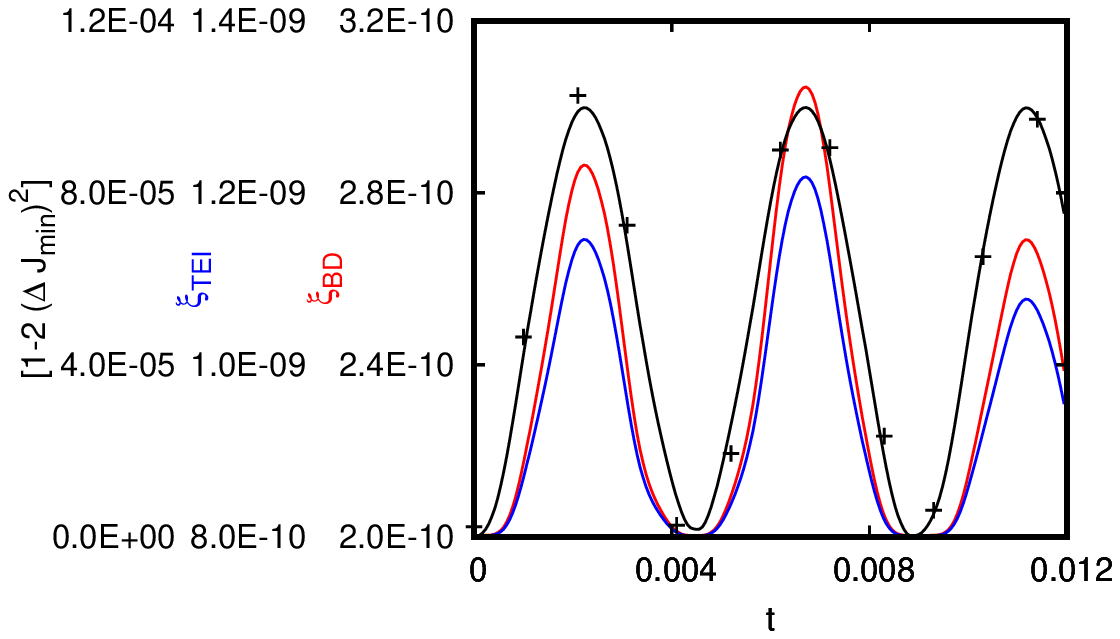}
%\xincludegraphics[width=0.31\textwidth,label=(c),fontsize=\scriptsize]{bipartite_freezing_BD.eps}
\caption{(a) $D( {}^{31}\mathrm{P} : {}^{19}\mathrm{F})$ (black), $\xi_{\smtextsc{qmi}}$ (blue), and $\xi_{\smtextsc{ipr}}$ (red) vs. time $t$ in seconds, and (b) $[1 - 2 (\Delta J_{\mathrm{min}})^{2}]$ (black), $\xi_{\smtextsc{tei}}$ (blue), and $\xi_{\smtextsc{bd}}$ (red) vs. time $t$ in seconds for Case (ii). The solid curves are computed by numerical simulation and the crosses from experimental data.}
\label{fig:bipart_freeze_1}
\end{figure*}

At each instant considered, the extent of spin squeezing $[1 - 2 (\Delta J_{\mathrm{min}})^{2}]$ is computed using the procedure described 
in Sec.~\ref{sec:squeeze_prop}. In this case, the mean spin direction $\mathbf{v}_{\smtextsc{s}}(t)$ is not a null vector. A set of unit vectors $\mathbf{v}_{\perp}$ that are orthogonal to $\mathbf{v}_{\smtextsc{s}}(t)$ was  obtained using a numerical search program. From these, we have computed the  variance and plotted $[1 - 2 (\Delta J_{\mathrm{min}})^{2}]$ as a function of time.
Figures \ref{fig:bipart_block} and \ref{fig:bipart_freeze_1} compare the indicators $\xi_{\smtextsc{tei}}$, $\xi_{\smtextsc{ipr}}$ and $\xi_{\smtextsc{bd}}$ with the discord, $[1 - 2 (\Delta J_{\mathrm{min}})^{2}]$, and $\xi_{\smtextsc{qmi}}$ for Cases (i) and (ii) respectively. From the plots, we infer that  all the numerically simulated tomographic indicators are in good agreement with the discord. In contrast to NMR experiment I, spin squeezing agrees well with the discord in experiment II. Further, the variance computed from the experimentally reconstructed density matrices and the numerically simulated density matrices are in good agreement. However, the tomographic indicators computed from the experiment do not match well with those from numerical simulations. This is especially the case when the purity of the state is very low ($\epsilon=10^{-4}$) and, consequently, the extent of entanglement is much  less than unity. We have also verified that the inferences drawn from Case (iii) are identical to that of Case (ii).

As was done in the case of the NMR experiment I, we have calculated $\xi^{\prime}_{\smtextsc{tei}}$ with different choices of a reduced number of tomographic slices. In contrast to our inference in NMR experiment I, none of the indicators ($\xi_{\smtextsc{tei}}$, $\xi_{\smtextsc{ipr}}$, and $\xi_{\smtextsc{bd}}$) are effective when computed from a reduced number of tomographic slices. The crucial difference between the two experiments is in the extent of entanglement. Typically, in NMR experiment I, discord is in the range $[0,1]$. Whereas in NMR experiment II, it is considerably less (the typical range is $[0,10^{-8}]$). It appears, therefore,  that $\xi^{\prime}_{\smtextsc{tei}}$ is a reliable entanglement indicator only for sufficiently strong bipartite entanglement. We corroborate this below, with further investigations pertaining to NMR experiment III.
	
\subsection{NMR experiment III}

The three qubit experiment has been performed on three nuclear spins by using $\vphantom{}^{13}\mathrm{C}$, $\vphantom{}^{1}\mathrm{H}$ and $\vphantom{}^{19}\mathrm{F}$ spins in DBFM (Fig.~\ref{fig:molecule} (a)), the same system as NMR experiment I described in Sec.~\ref{sec:NMRexpt}. Here $\vphantom{}^{13}\mathrm{C}$, $\vphantom{}^{1}\mathrm{H}$, and $\vphantom{}^{19}\mathrm{F}$ are subsystems $1,\,2,$ and $3$ respectively. The Hamiltonian of the system is of the form
\begin{equation}
H_{\smtextsc{iii}}= \sum\limits_{i=1}^{3} \omega_{i} \sigma_{i x} -\sum\limits_{i=1}^{3} \Omega_i \sigma_{i z} + H_{\smtextsc{i}}
\label{eqn:Hnmr3qRyd}
\end{equation}
where $H_{\smtextsc{i}}$ is as defined in Eq.~\eref{eqn:HJ1} (rewritten in terms of subscripts 1, 2 and 3 instead of M, A and B respectively). It can be seen that Eq.~\eref{eqn:Hnmr3qRyd} can be obtained by re-writing Eq.~\eref{eqn:Ham_nmr2} appropriately. The initial density matrix given in Eq.~\eref{eqn:rhoNinit} was numerically evolved in time under the Hamiltonian given by Eq.~\eref{eqn:Ham_nmr2} for $N=3$, and the discord computed as a function of time~\cite{nmrExpt1}, for $\lambda_{12}/(2 \pi) = 224.7$ Hz, $\lambda_{13}/(2 \pi) =-311.1$ Hz, $\lambda_{23}/(2 \pi) = 49.7$ Hz, $\Omega_{1}=(\lambda_{12}+\lambda_{13})/2$, $\Omega_{2}=(\lambda_{12}+\lambda_{23})/2$ and $\Omega_{3}=(\lambda_{13}+\lambda_{23})/2$. Four cases have been examined, namely, (A) $\omega_{1}/(2 \pi)=10$ Hz, $\omega_{1}=\omega_{2}=\omega_{3}$ [bipartite entanglement between subsystems (1) and (2,3)], (B) $\omega_{1}/(2 \pi)=50$ Hz, $\omega_{1}=5 \omega_{2} = 5 \omega_{3}$ [bipartite entanglement between subsystems (1) and (2,3)], (C) $\omega_{2}/(2 \pi)=50$ Hz, $\omega_{2}=5 \omega_{1}=5 \omega_{3}$ [bipartite entanglement between subsystems (2) and (1,3)], and (D) $\omega_{1}/(2 \pi)=50$ Hz, $\omega_{1}=\omega_{2}=5 \omega_{3}$ [bipartite entanglement between subsystems (1,2) and (3)]. Using a similar procedure, we have now computed the corresponding $\xi_{\smtextsc{qmi}}$. The tomographic indicators $\xi_{\smtextsc{tei}}$, $\xi_{\smtextsc{ipr}}$, $\xi_{\smtextsc{bd}}$ and spin squeezing $[1 - (4/3) (\Delta J_{\mathrm{min}})^{2}]$ were computed, as before.
In this case, the spin operator 
\begin{align*}
\mathbf{J}_{3}= &(\sigma_{1 x} + \sigma_{2 x} + \sigma_{3 x}) \mathbf{e_{x}} + (\sigma_{1 y} + \sigma_{2 y} + \sigma_{3 y}) \mathbf{e_{y}} \\
&+ (\sigma_{1 z} + \sigma_{2 z} + \sigma_{3 z}) \mathbf{e_{z}},
\end{align*}
is used instead of $\mathbf{J}_{2}$ defined in Eq.~\eref{eqn:spinJ}. Here $(\Delta J_{\mathrm{min}})^{2}<0.75$ implies spin squeezing.

We have verified that the plots pertaining to the Cases (A) and (B) are similar to the corresponding plots in Figs. \ref{fig:bipart_block} and \ref{fig:bipart_freeze_1} respectively. 
As in NMR experiments I and II, in Cases (A) and (B) of experiment III, the gross features of entanglement are captured by the indicators. In contrast to experiment I, and similar to experiment II, spin squeezing agrees with discord in Cases (A) and (B). The latter, when computed from the experimentally reconstructed density matrices and from numerical simulations,  agree well with each other. 
	
\begin{figure*}[t]
\xincludegraphics[width=0.45\textwidth,label=(a),fontsize=\scriptsize]{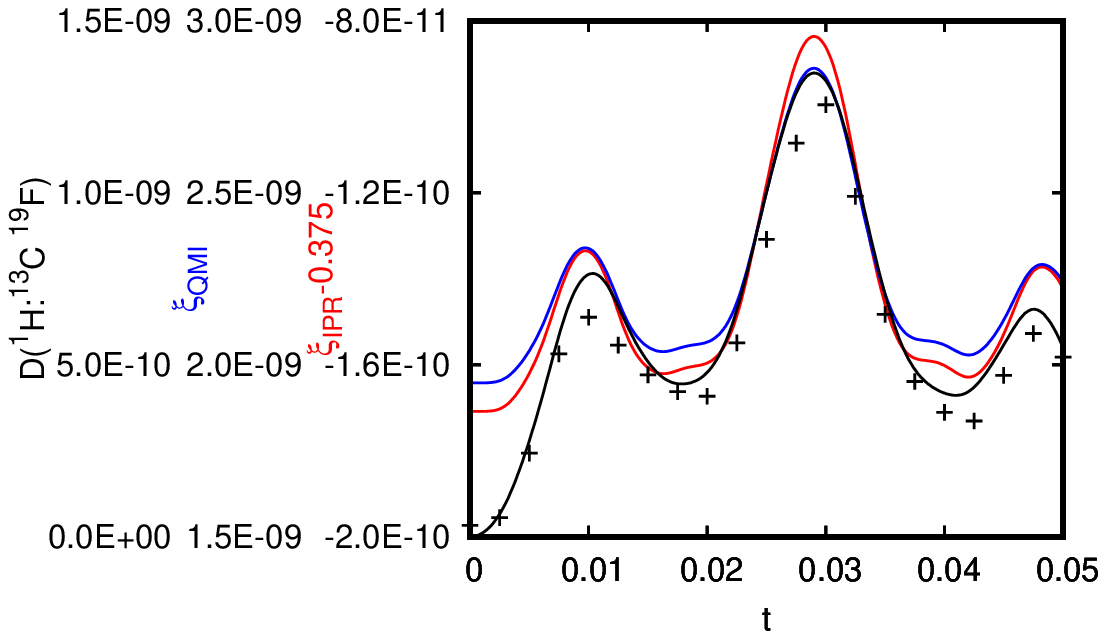}
\xincludegraphics[width=0.45\textwidth,label=(b),fontsize=\scriptsize]{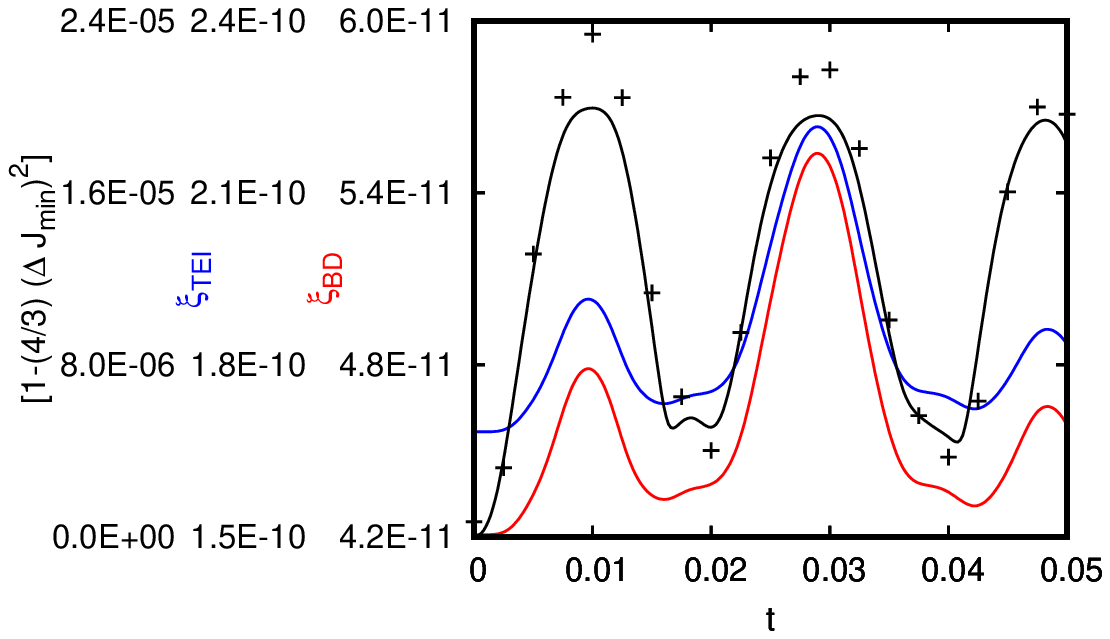}
%\xincludegraphics[width=0.31\textwidth,label=(c),fontsize=\scriptsize]{tripartite_freezing_1_BD.eps}
\caption{
(a) $D( {}^{1}\mathrm{H} : {}^{13}\mathrm{C} \: \: {}^{19}\mathrm{F})$ (black), $\xi_{\smtextsc{qmi}}$ (blue), and $\xi_{\smtextsc{ipr}}$ (red) vs. time $t$ in seconds, and (b) $[1 - (4/3) (\Delta J_{\mathrm{min}})^{2}]$ (black), $\xi_{\smtextsc{tei}}$ (blue), and $\xi_{\smtextsc{bd}}$ (red) vs. time $t$ in seconds for Case (C). The solid curves are computed by numerical simulation and the crosses from experimental data.
}
\label{fig:tripart_freeze_2}
\end{figure*}

\begin{figure*}[t]
\xincludegraphics[width=0.45\textwidth,label=(a),fontsize=\scriptsize]{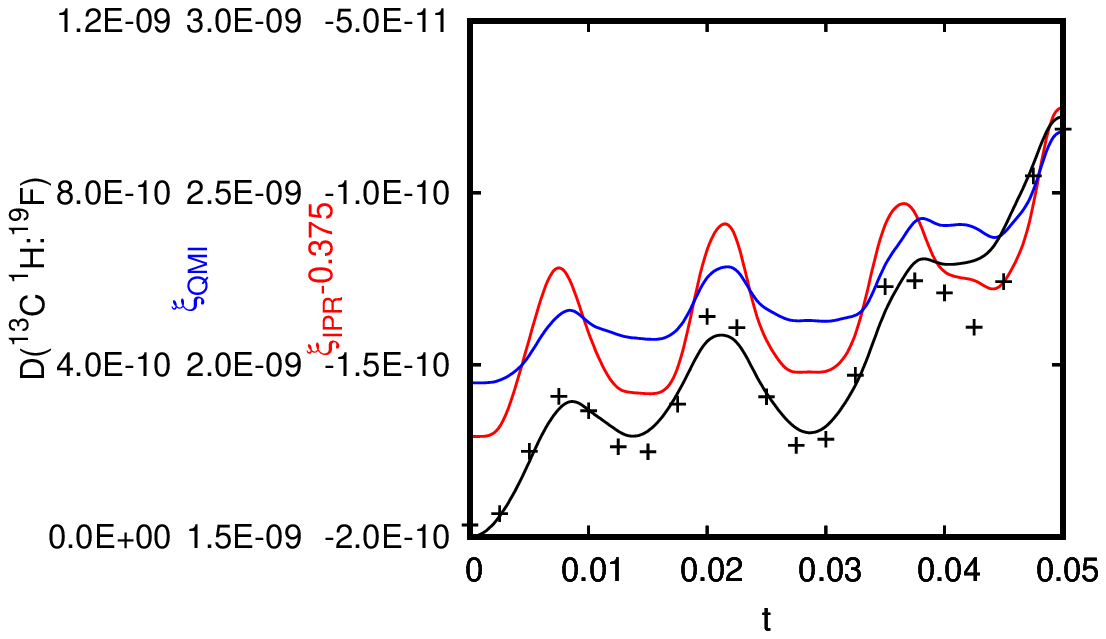}
\xincludegraphics[width=0.45\textwidth,label=(b),fontsize=\scriptsize]{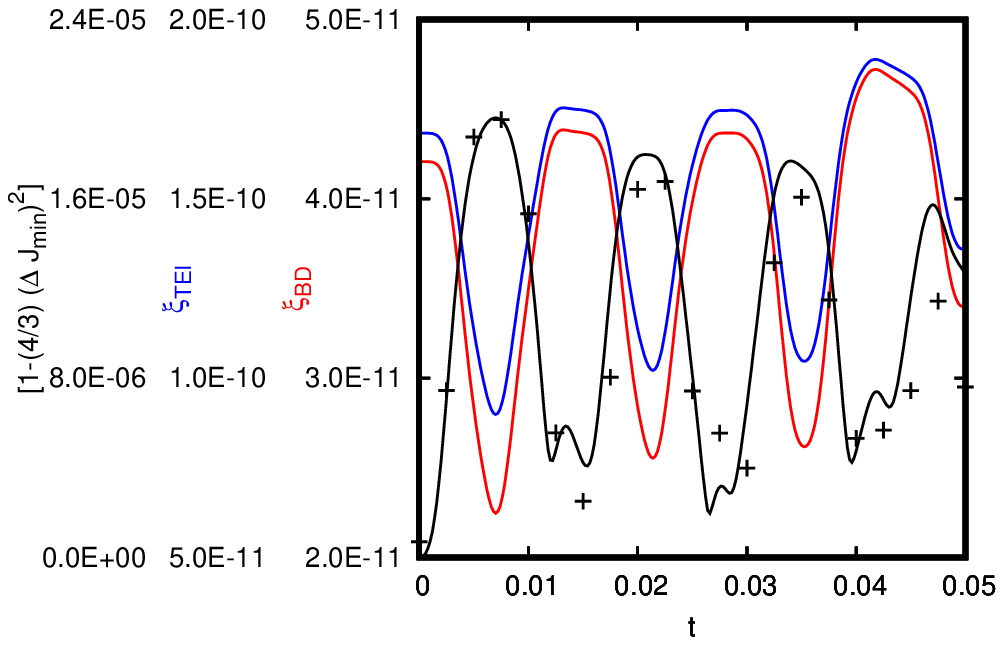}
%\xincludegraphics[width=0.31\textwidth,label=(c),fontsize=\scriptsize]{tripartite_freezing_2_BD.eps}
\caption{
(a) $D( {}^{13}\mathrm{C} \: \: {}^{1}\mathrm{H} : {}^{19}\mathrm{F})$ (black), $\xi_{\smtextsc{qmi}}$ (blue), and $\xi_{\smtextsc{ipr}}$ (red) vs. time $t$ in seconds, and (b) $[1 - (4/3) (\Delta J_{\mathrm{min}})^{2}]$ (black), $\xi_{\smtextsc{tei}}$ (blue), and $\xi_{\smtextsc{bd}}$ (red) vs. time $t$ in seconds for Case (D). The solid curves are computed by numerical simulation and the crosses from experimental data.
}
\label{fig:tripart_freeze_3}
\end{figure*}

Figures \ref{fig:tripart_freeze_2} and \ref{fig:tripart_freeze_3} compare the indicators $\xi_{\smtextsc{tei}}$, $\xi_{\smtextsc{ipr}}$ and $\xi_{\smtextsc{bd}}$ with the discord, $[1 - (4/3) (\Delta J_{\mathrm{min}})^{2}]$, and $\xi_{\smtextsc{qmi}}$  for Cases (C) and (D) respectively. In Case (C), all tomographic indicators agree with the discord as before. However, as in experiment I, and in contrast with the other cases that have been examined till now, $[1 - (4/3) (\Delta J_{\mathrm{min}})^{2}]$ does not mimic the discord effectively. 

The inferences in Case (D) are in stark contrast to those obtained till now. The unanticipated result is that $\xi_{\smtextsc{tei}}$, $\xi_{\smtextsc{bd}}$ and spin squeezing do not mirror the temporal behaviour of the discord well. Only $\xi_{\smtextsc{ipr}}$ agrees reasonably well with the discord.

We note that, as in NMR experiment II, here too  the bipartite entanglement is weak, and reducing the number of tomographic slices is not effective. We emphasize that the plots in Figs. \ref{fig:tripart_freeze_2} and \ref{fig:tripart_freeze_3} have been obtained retaining all the 27 tomographic slices.

\section{Discussion and conclusion}

Earlier work~\cite{sharmila4} indicates that $\xi_{\smtextsc{pcc}}$, which captures linear correlations between the respective quadratures of the two subsystems, does not reliably reflect the extent of entanglement. In direct contrast to this, in NMR experiment I where the state considered is a \textit{mixed} bipartite state, $\xi_{\smtextsc{pcc}}$ is in good agreement with spin squeezing and negativity $N(\rho_{\smtextsc{ab}})$, although not with the discord. Results from both the NMR experiments I and II indicate that all other tomographic entanglement indicators are in good agreement with the discord. We note that NMR experiment II deals with blockade and freezing of spins. We have further shown that novel features could arise in this regard when a tripartite system is partitioned into two subsystems and bipartite entanglement examined, as is revealed by Case (D) of NMR experiment III. Thus, the performance of entanglement indicators, and the extent to which these indicators and spin squeezing track  the discord during dynamical evolution of multipartite systems, are very sensitive to the precise manner in which  the full system 
is partitioned into subsystems, as well as to features such as blockade and spin freezing.

Our investigations on these three systems provide some preliminary pointers on the efficacy of identifying an optimal subset of tomographic slices for the computation of the corresponding bipartite entanglement indicator $\xi^{\prime}_{\smtextsc{tei}}$. If the entanglement is sufficiently strong (NMR experiment I), a subset of tomographic slices suffice. For the experiment, this would indicate a corresponding reduced number of measurements. However, if the entanglement is weak (NMR experiments II and III), the full set of tomographic slices need to be used. A relevant question is whether the closeness of $\xi^{\prime}_{\smtextsc{tei}}$ to $\xi_{\smtextsc{tei}}$ implies strong bipartite entanglement. More detailed investigations need to be carried out  before this question can be  answered.

\backmatter

\bmhead{Acknowledgments}

This work was supported in part by a seed grant from IIT Madras to C-QuICC (Centre for Quantum Information, Communication and Computing) under the IoE-CoE scheme.

\begin{appendices}
\section*{Appendix: Equivalent circuit for NMR experiment I}
\setcounter{section}{1}

\begin{figure*}[t]
\centering
\xincludegraphics[width=0.5\textwidth,label=(a),fontsize=\scriptsize]{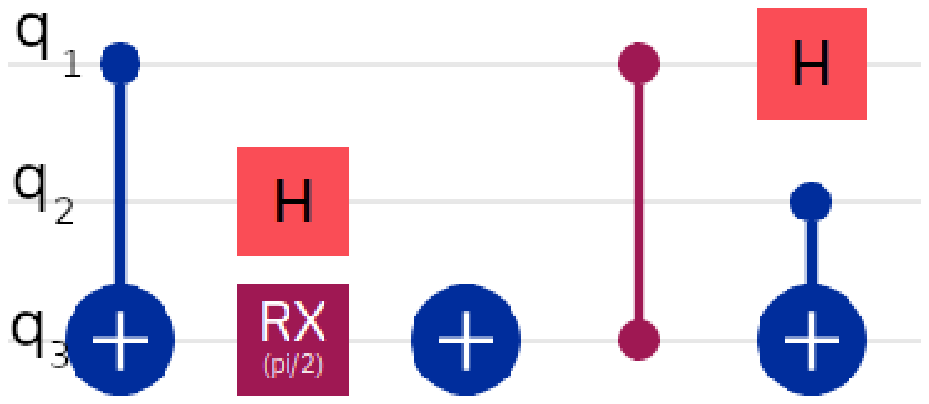}
\hspace*{1. em}
\xincludegraphics[width=0.24\textwidth,label=(b),fontsize=\scriptsize]{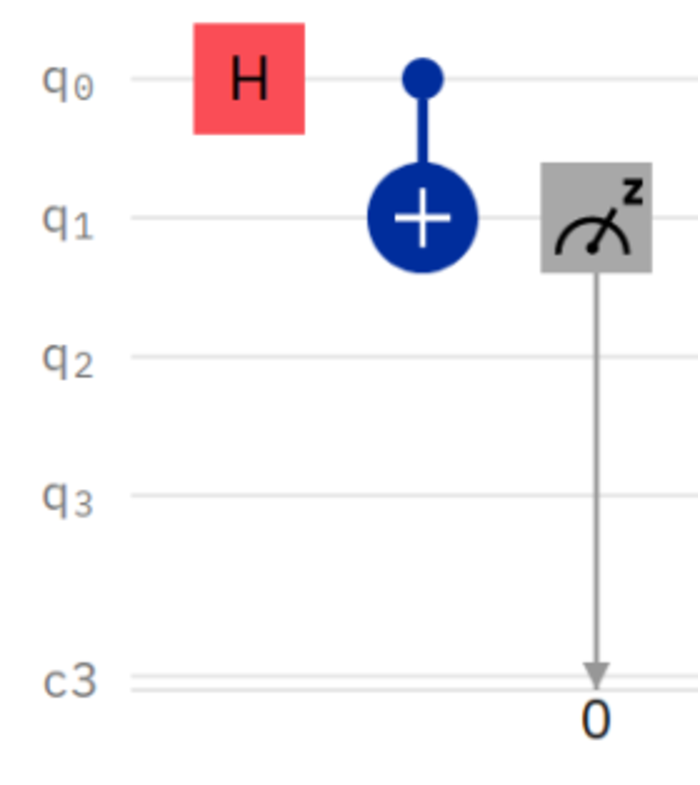}
\caption{Parts of the equivalent circuit for NMR experiment I (created using IBM Q).}
\label{fig:circuit_parts}
\end{figure*}

Corresponding to any instant, the equivalent circuit comprises two parts. Without loss of generality, we describe the circuit at instant $\chi_{\mathrm{s}}t=\pi/8$ (Figs.~\ref{fig:circuit_parts}(a) and (b)). Here,
$q_{1}$, $q_{2}$ and $q_{3}$ are the qubits that follow the dynamics of the subsystems M, A, and B respectively, and,
\begin{equation}
RX(\theta)= \begin{bmatrix}
\cos\,(\theta/2) &\, -i\,\sin\,(\theta/2) \\
-i\,\sin\,(\theta/2)&\, \cos\,(\theta/2) 
\end{bmatrix},
\label{eqn:RX}
\end{equation} 
with  $0\leqslant\theta<\pi$. We note that $\theta$ is analogous to $4 \chi_{\mathrm{s}} t$ in the NMR experiment, and $\chi_{\mathrm{s}} t=\pi/8$ corresponds to $\theta=\pi/2$. 
The circuit in Fig.~\ref{fig:circuit_parts}(a) performs the following task. If the initial state of $q_{1}$, $q_{2}$, and $q_{3}$ is set to $\ket{0}$, the final state is analogous to $\ket{\Psi_{0}}$ (defined in Eq.~\eref{eqn:Psi1}). If the initial states of the three qubits are respectively $\ket{1}$, $\ket{0}$, and $\ket{0}$, then the output state is analogous to $\ket{\Psi_{1}}$ (also defined in Eq.~\eref{eqn:Psi1}). 
	
\begin{figure*}[t]
\centering
\includegraphics[width=0.8\textwidth]{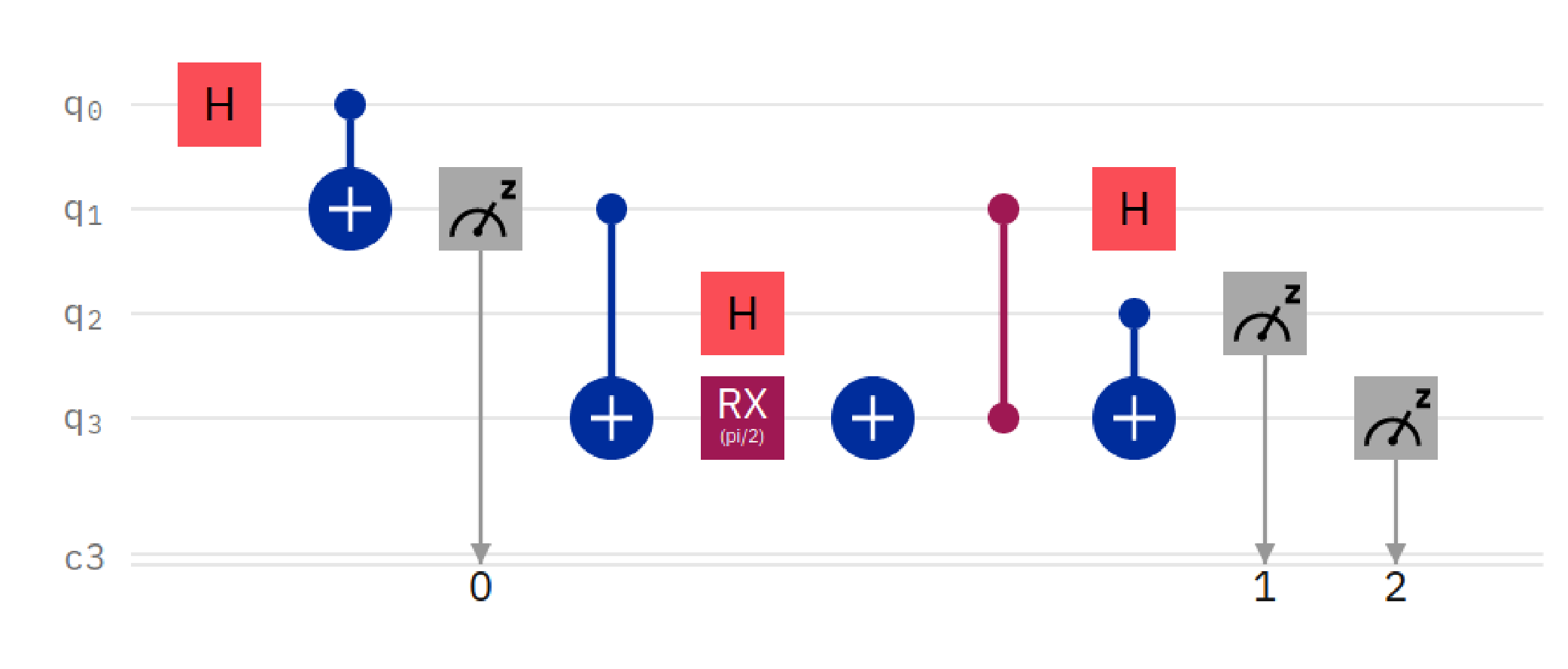}
\caption{Equivalent circuit for NMR experiment I (created using IBM Q).}
\label{fig:circuit_dia}
\end{figure*}

Thus, by setting the initial state of $q_{1}$ to be $\frac{1}{2} \left(\ket{0} \bra{0} + \ket{1} \bra{1}\right)$, we can obtain the desired output state analogous to $\rho_{\smtextsc{mab}}(t)$. This mixed state is achieved in Fig.~\ref{fig:circuit_parts}(b) using an auxiliary qubit $q_{0}$. Tomograms corresponding to $\rho_{\smtextsc{ab}}(t)$ have been obtained from measurement outcomes on $q_{2}$ and $q_{3}$ (Fig.~\ref{fig:circuit_dia}).

\begin{figure*}[t]
\centering
\xincludegraphics[width=0.3\textwidth,label=(a),fontsize=\scriptsize]{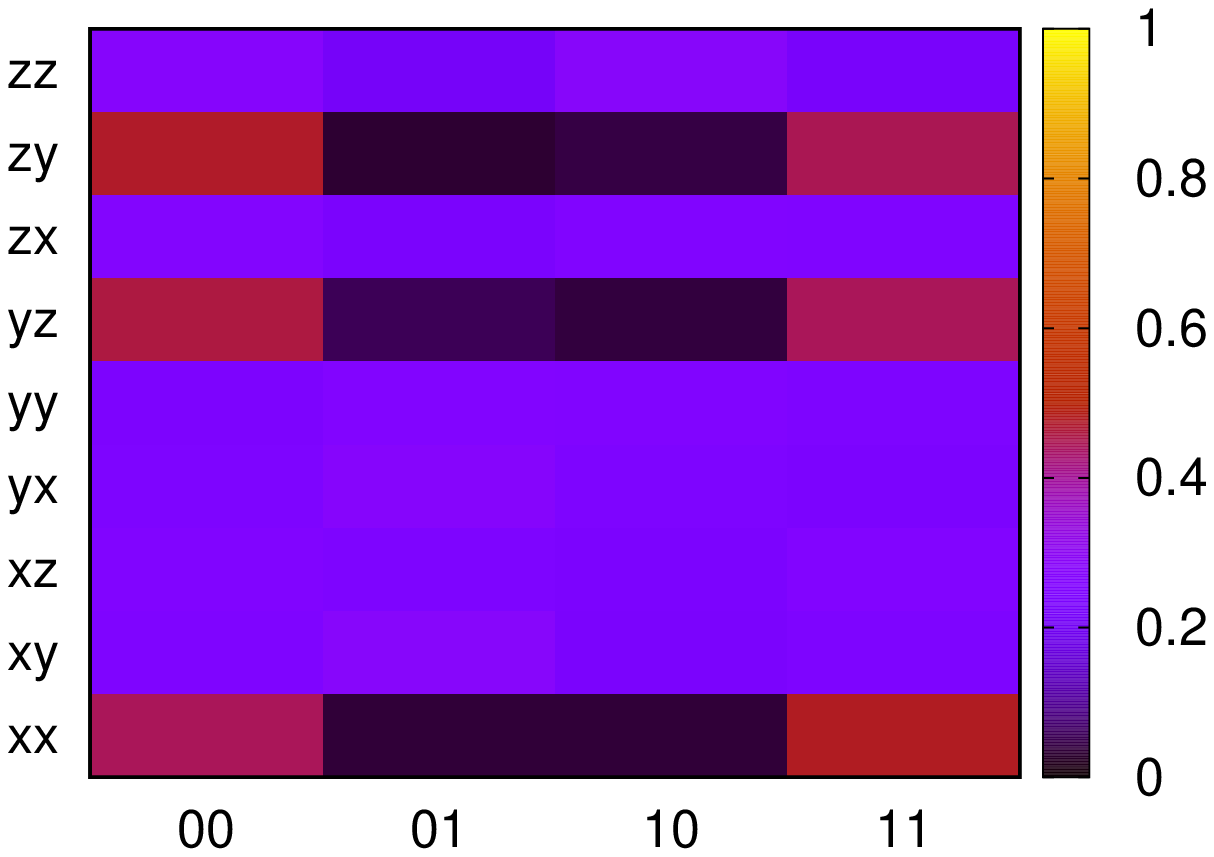}
\xincludegraphics[width=0.3\textwidth,label=(b),fontsize=\scriptsize]{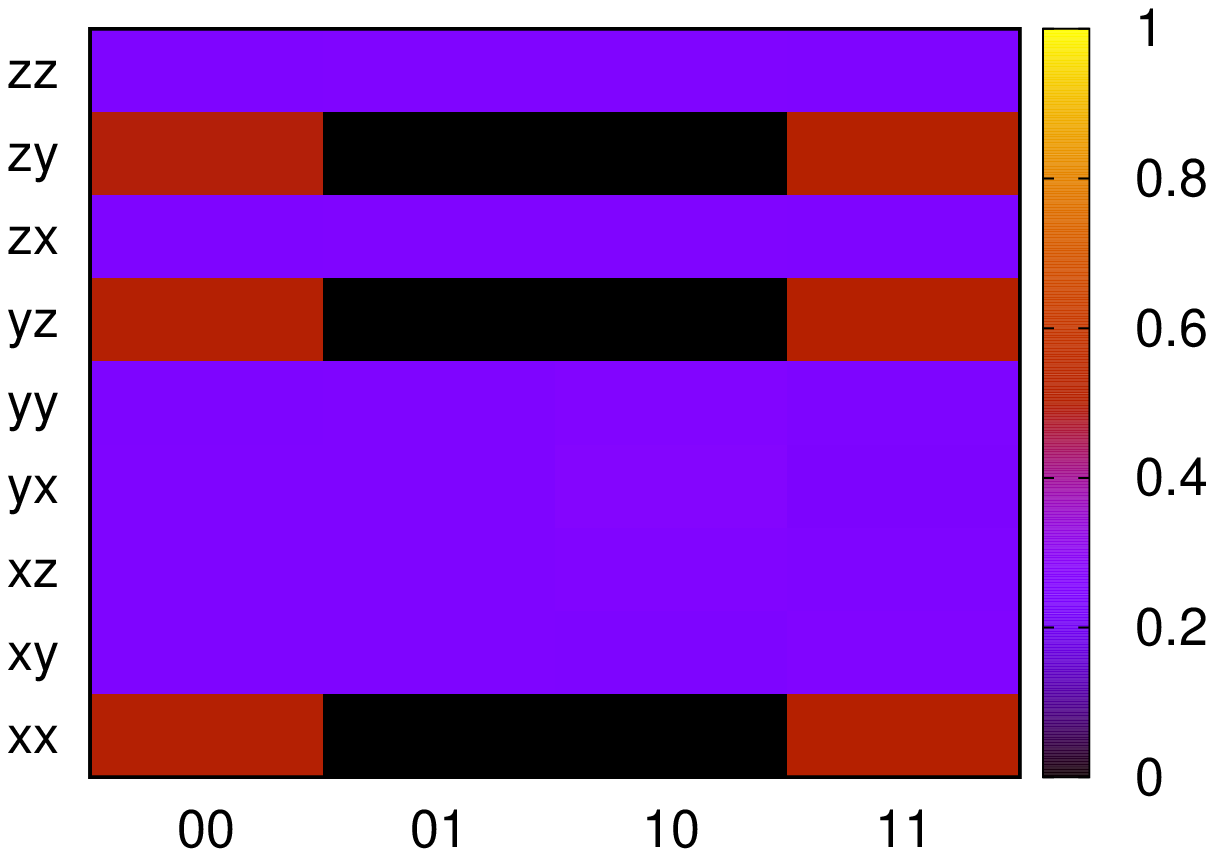}
\xincludegraphics[width=0.3\textwidth,label=(c),fontsize=\scriptsize]{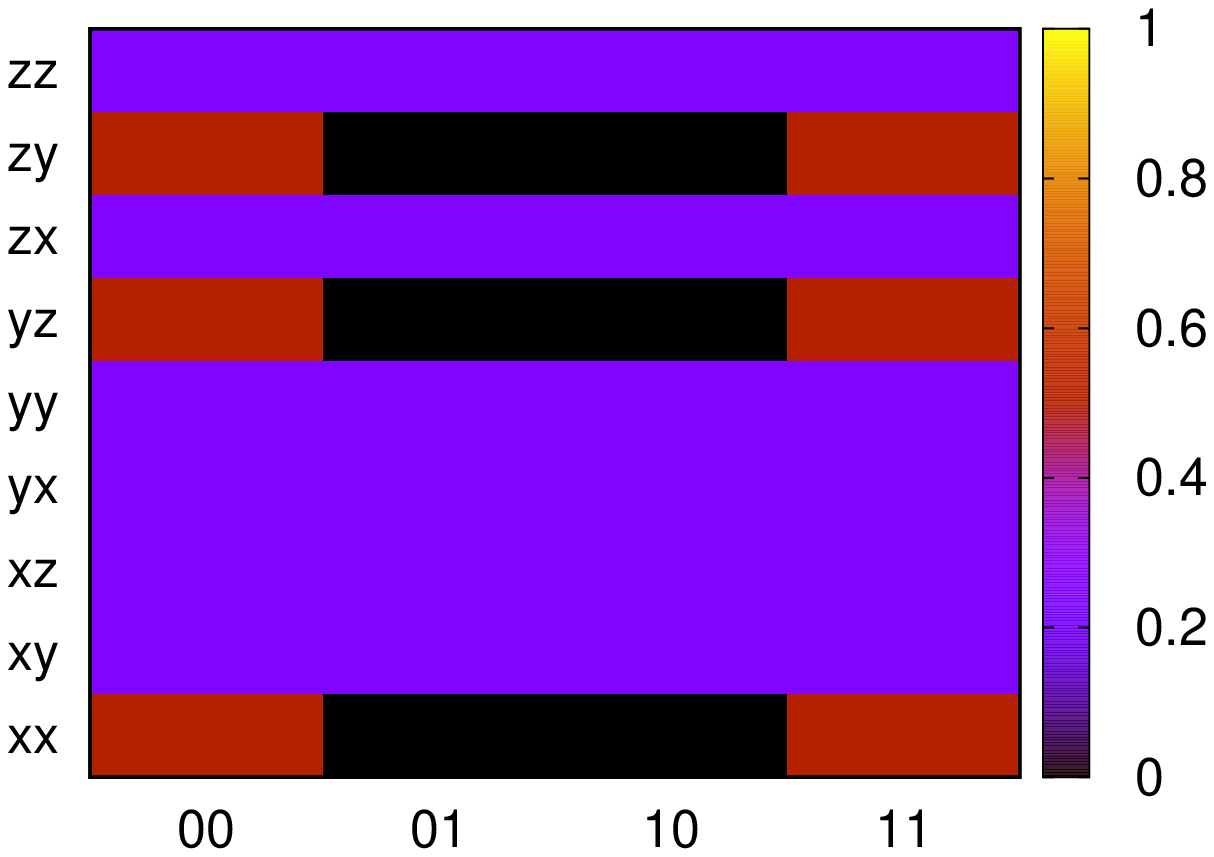}
\caption{Tomograms from (a) IBM Q experiment (b) QASM simulation (c) numerical computations of NMR experiment I.}
\label{fig:tomograms}
\end{figure*}

Measurements are carried out in the $x$, $y$ and $z$ bases corresponding to the matrices defined in Eq.~\eref{eqn:atomops}. A measurement in the $z$-basis is automatically provided by the IBM platform. A measurement in the $x$-basis is achieved by applying an Hadamard gate followed by a $z$-basis measurement.  Defining the operator
\begin{equation}
S^{\dagger} = \begin{bmatrix}
1 & 0 \\
0 & -i \end{bmatrix},
\label{eqn:sdg}
\end{equation}
measurement in the $y$-basis is achieved by applying $S^{\dagger}$, then  an Hadamard gate, and  finally a measurement in 
the $z$-basis. Measurements in the  $x$, $y$ and $z$ bases  are needed for obtaining the spin tomogram in Fig.~\ref{fig:tomograms} (a). (This is equivalent to the bipartite spin tomogram in NMR experiment I at $\chi_{\mathrm{s}}t=\pi/8$, in the basis states corresponding to $\sigma_{x}$, $\sigma_{y}$ and $\sigma_{z}$). 

In the NMR setup, complete state tomography is possible by rotating the qubits via appropriate unitary transformations, and performing measurements in one fixed convenient basis~\cite{NMRtomography}. Similarly, in the IBM quantum computer, the convenient basis for measurement was the $z$-basis, as it is the one naturally available in the platform. To measure in any other basis, we resort to rotating the qubit using appropriate gates. Hence, the two measurement processes are comparable.

The spin tomograms have also been obtained  experimentally 
using the IBM superconducting circuit with appropriate Josephson junctions  (Fig.~\ref{fig:tomograms} (a)), and the QASM simulator  provided by IBM. The latter does not take into account losses at various stages of the circuit (Fig.~\ref{fig:tomograms} (b)). These tomograms have been compared with the numerically generated nuclear spin tomogram where decoherence effects have been neglected (Fig.~\ref{fig:tomograms} (c)).
	
Six executions of the experiment were carried out. Each execution of the experiment comprised $8192$ runs over each of the 9 basis sets. From these six tomograms, $\xi_{\smtextsc{tei}}$ has been calculated. The values 
obtained from the experiment, simulation and numerical analysis 
are $0.1941\pm0.0083$, $0.3334$ and $0.3333$,  respectively. The error bar in the experimental value
was calculated from the standard deviation of $\xi_{\smtextsc{tei}}$.
The experimental value differs from those obtained in the simulation of circuit and the model, because of inevitable losses. 

\begin{figure*}[t]
\centering
\includegraphics[width=0.8\textwidth]{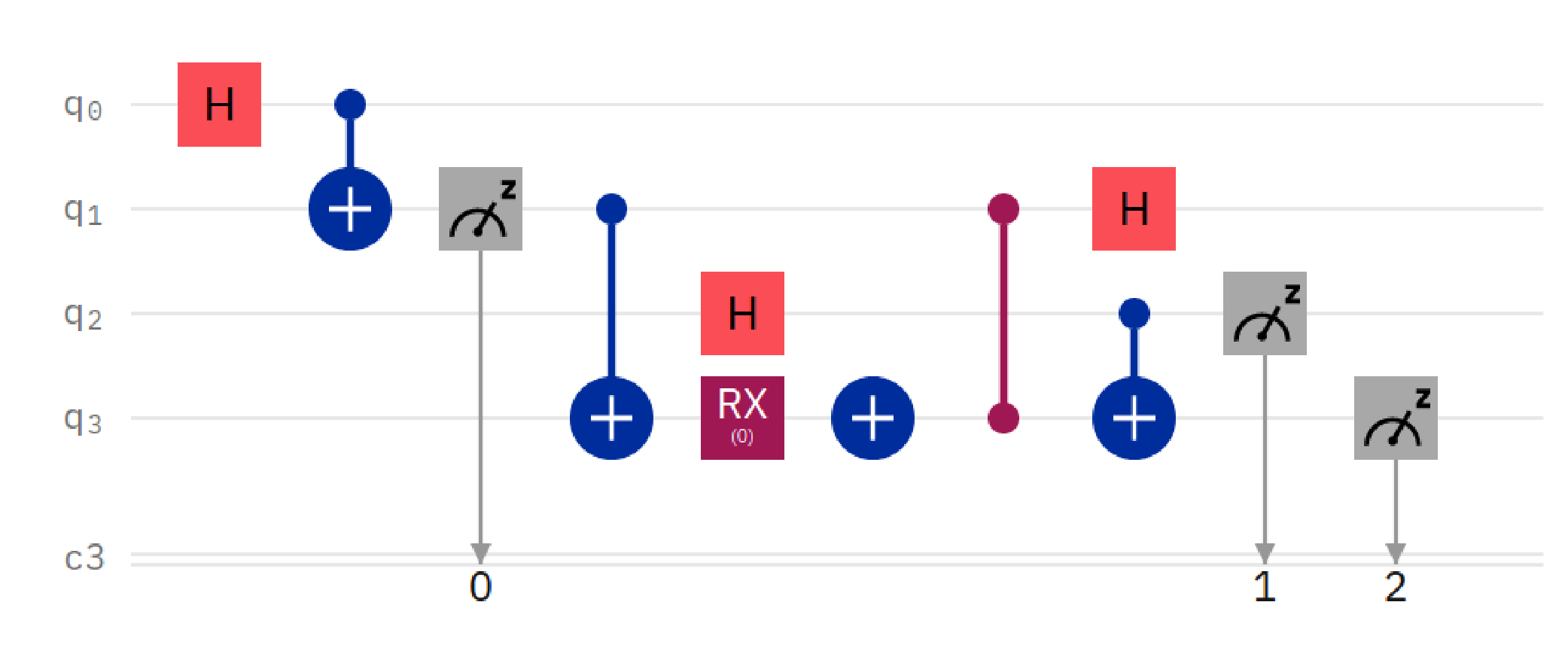}
\caption{Equivalent circuit for initial state analogous to $\rho_{\smtextsc{mab}}(0)$ in NMR experiment I (created using IBM Q).}
\label{fig:init_circuit}
\end{figure*}

\begin{figure*}[t]
\centering
\includegraphics[width=0.8\textwidth]{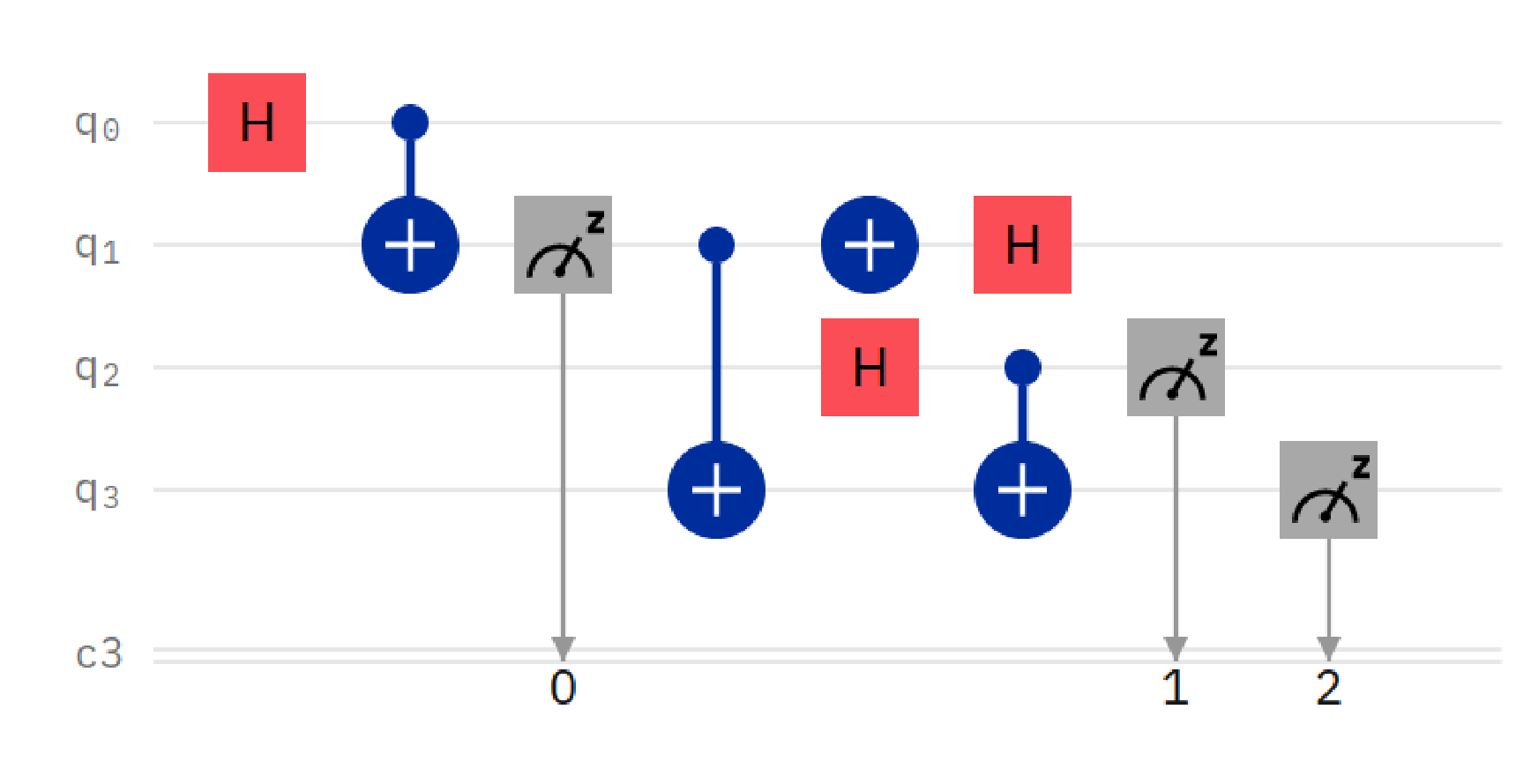}
\caption{An alternative equivalent circuit for initial state analogous to $\rho_{\smtextsc{ab}}(0)$ in NMR experiment I (created using IBM Q).}
\label{fig:init_circ_2}
\end{figure*}

The circuit corresponding to the initial state can be obtained by setting $\theta$ to zero. The corresponding values of $\xi_{\smtextsc{tei}}$ obtained from the experiment, simulation of circuit and model are $0.0676\pm0.0065$, $0.1113$ and $0.1111$ respectively. This demonstrates that substantial losses arise even in generating the initial state using this circuit. It is more efficient to construct the initial state circuit in \cite{nmrExpt} (Fig.~\ref{fig:init_circ_2}) as it involves a smaller  number of components.

The values for $\xi_{\smtextsc{tei}}$ obtained from the experiment, simulation and model corresponding to this circuit are $0.0720\pm0.0025$, $0.1112$ and $0.1111$ respectively. As expected, the experimental value agrees better with the simulation of circuit and numerical computation of the model.

\end{appendices}

\bibliography{references}

\end{document}